\documentclass[onecolumn,amsmath,amssymb,12pt,superscriptaddress,nofootinbib]{revtex4-1}
\pdfoutput=1

\usepackage[english]{babel}
\usepackage{amsthm}
\usepackage{amssymb}
\usepackage{amsmath}
\usepackage{amsthm}
\usepackage[]{graphicx}
\usepackage{tensor}
\usepackage{color}
\usepackage{framed}
\usepackage{framed}
\usepackage[bookmarks,linktocpage, colorlinks=true, plainpages = false, citecolor = blue,  linkcolor=blue, urlcolor = blue, filecolor = blue]{hyperref} 
\usepackage{natbib}
\usepackage{dcolumn}
\usepackage{bm}

\usepackage{tikz}
\usetikzlibrary{decorations.pathmorphing, patterns,shapes}

\input epsf

\begin{document}

\allowdisplaybreaks
\begin{titlepage}

\title{
Quantum Incompleteness of Inflation
}
\author{Alice Di Tucci}
\email{alice.di-tucci@aei.mpg.de}
\affiliation{Max--Planck--Institute for Gravitational Physics (Albert--Einstein--Institute), 14476 Potsdam, Germany}
\author{Job Feldbrugge}
\email{jfeldbrugge@perimeterinstitute.ca}
\affiliation{Perimeter Institute, 31 Caroline St N, Ontario, Canada}
\author{Jean-Luc Lehners}
\email{jlehners@aei.mpg.de}
\affiliation{Max--Planck--Institute for Gravitational Physics (Albert--Einstein--Institute), 14476 Potsdam, Germany}
\author{Neil Turok}
\email{nturok@perimeterinstitute.ca}
\affiliation{Perimeter Institute, 31 Caroline St N, Ontario, Canada}
\begin{abstract}
\vspace{.5cm}
\noindent 
Inflation is most often described using quantum field theory (QFT) on a fixed, curved spacetime background. Such a description is valid only if the spatial volume of the region considered is so large that its size and shape moduli behave classically. However, if we trace an inflating universe back to early times, the volume of any comoving region of interest -- for example the present Hubble volume -- becomes exponentially small. Hence, quantum fluctuations in the trajectory of the background cannot be neglected at early times. In this paper, we develop a path integral description of a flat, inflating patch (approximated as de Sitter spacetime), treating both the background scale factor and the gravitational wave perturbations quantum mechanically. We find this description fails at small values of the initial scale factor, because \emph{two} background saddle point solutions contribute to the path integral. This leads to a breakdown of QFT in curved spacetime, causing the fluctuations to be unstable and out of control. We show the problem may be alleviated by a careful choice of quantum initial conditions, for the background and the fluctuations, provided that the volume of the initial, inflating patch is much larger than $H^{-1}$ in Planck units with $H$ the Hubble constant at the start of inflation. The price of the remedy is high: not only the inflating background, but also the stable, Bunch-Davies fluctuations must be input by hand. Our discussion emphasizes that, even if the inflationary scale is far below the Planck mass, new physics is required to explain the initial quantum state of the universe.
\end{abstract}
\maketitle
\end{titlepage}
\tableofcontents
\section{Introduction}

One often thinks of gravity as describing the very large, and quantum mechanics the very small. However, one of the most stunning ideas to emerge from contemporary cosmology is that the largest visible structures in the universe originated through the amplification of primordial quantum fluctuations. It is frequently argued that inflation \cite{Guth:1980zm,Linde:1981mu,Albrecht:1982wi} provides a natural mechanism to achieve this amplification, in a rather generic manner. The calculations underpinning this assertion are performed in the framework of quantum field theory (QFT) in curved spacetime, where the background spacetime is treated classically and only the fluctuations are quantized \cite{Mukhanov:1981xt,Starobinsky:1979ty,Starobinsky:1982ee, Guth:1982ec,Hawking:1982cz,Bardeen:1983qw}. There is an implicit assumption that this classical/quantum split between background and perturbations is a good approximation to a more fundamental theory of quantum gravity, where the complete four-geometry is treated quantum mechanically. In the present work we test this assumption by going beyond QFT in curved spacetime and studying inflationary quantum dynamics using the path integral for Einstein gravity within a semi-classical expansion. 

There is a fundamental difference in the behavior of quantum and the classical relativistic and diffeomorphism-invariant theories, which lies at the heart of our work. A classical relativistic particle, for example, cannot change the sign of the time component of its four-velocity. If it is travelling forwards in time at one moment, it will always travel forwards. However, a quantum relativistic particle cannot be so constrained: it can ``turn around'' in time. Indeed, including such amplitudes is essential to the final consistency of the theory (for a nice discussion see, for example, Feynman's Dirac Memorial lecture ~\cite{Feynman:1987gs}). 

The scale factor of the universe, being a time-like coordinate on superspace, is analogous to the time coordinate for a particle. The usual, classical picture of inflation is of a universe which always expands. However, when we study the quantum dynamics of the cosmological background, we have no right to exclude trajectories for which the scale factor turns around, {\it i.e}, they may start out contracting before expanding to attain their very large, final size. Even in classical general relativity, there is the example of de Sitter spacetime in the closed (global) slicing.  If one fixes the initial and the final radius to be both greater than the de Sitter radius, then there evidently exist {\it two} classical saddle point solutions to the gravitational path integral, one of which undergoes a ``bounce.'' It seems self-evident that {\it both} such saddle point solutions  must be included in the transition amplitude between the initial and final three-geometries. 

In the case of the flat slicing of de Sitter (where we take a toroidal spatial universe in order to keep the action finite), it turns out there are likewise two classical solutions, both of which are relevant saddle points for the gravitational path integral giving the amplitude for an initial small universe to become a large one. One of the solutions contracts to touch zero size before inflating to the final size. We shall show that this background comes along with perturbations which are unstable and out of control. Unfortunately, since we are doing quantum mechanics we cannot exclude this background solution. One might hope that by taking the size of the initial universe to zero, one could obtain a solution which only expanded ``from nothing.'' We shall carefully study this limit and show that the unstable perturbations are unavoidable, at least within semi-classical gravity. Likewise, we shall study the flat limit of a closed de Sitter universe. Again, we find no way to avoid these problematical perturbations. The only way out of the problem, as we explain in Section VI,  appears to be to choose an ``off-shell,'' localized initial state for the universe, so that the universe is, at the start, sufficiently large and expanding that contributions from initially contracting universes are suppressed.  

Our discussion extends recent studies of the no-boundary proposal~\cite{Hartle:1983ai}, based on the Lorentzian path integral for gravity~\cite{Feldbrugge:2017kzv,Feldbrugge:2017fcc,DiazDorronsoro:2017hti,Feldbrugge:2017mbc,DiazDorronsoro:2018wro,Feldbrugge:2018gin}, see also \cite{Kamenshchik:2018rpw}. In those works, we found an interesting interplay between the cosmological background and the perturbations, when both are treated quantum mechanically. In particular, we found that imposing a ``no boundary'' initial condition, {\it i.e.}, that the universe began on a three-geometry of zero size, results in unstable perturbations. In this paper we generalize these calculations to scenarios for the beginning of inflation. We show that if an initial inflationary patch is assumed to start out much smaller than the Hubble volume, but still much larger than the Planck volume, there are generally {\it two} relevant semiclassical backgrounds, one of which ``bounces.'' We then show that the quantum perturbations about the bouncing background are badly behaved and out of control. This finding is closely related to our recent demonstration that the path integral formulations of the  ``no boundary" proposal of Hartle and Hawking, as well as the ``tunneling" proposal of Vilenkin, as originally proposed, lead to unacceptable quantum fluctuations~\cite{Feldbrugge:2017kzv,Feldbrugge:2017fcc,Feldbrugge:2017mbc,Feldbrugge:2018gin}. Here we make the statement more general: if one assumes a (quasi-) de Sitter phase all the way back to the beginning of the universe, and attempts to describe this phase using semi-classical Einstein gravity, then one picks out the {\it unstable} fluctuation mode rather than the stable, Bunch-Davies mode. We conclude that, when both the background and the fluctuations are quantized, inflation does {\it not} automatically pick out the Bunch-Davies vacuum, with associated nearly Gaussian-distributed fluctuations.  At face value, our finding invalidates the usual predictions of inflationary models unless some additional mechanism is invoked to explain the ``initial'' Bunch-Davies vacuum state. In other words, inflation on its own cannot explain the origin of the primordial perturbations. It is in this sense that we conclude that inflation is quantum incomplete.

One may then wonder under what circumstances the usual treatment of inflation, using QFT in a classical curved background spacetime, may be recovered. We shall show that a standard QFT description may be recovered provided that suitable initial conditions are imposed. Namely, we assume a localized initial quantum state describing an {\it expanding} universe of a prescribed size. We propagate this state forward using the Feynman path integral propagator. Provided the assumed initial size (three-volume) of the universe is sufficiently large,  then indeed only one, monotonically expanding background solution is relevant to the path integral for gravity. For this calculation, it turns out to be crucial that we integrate over positive values of the lapse rather than both positive and negative values, as has been advocated by some. The reason is that the initial quantum state specifies the initial momentum conjugate to the scale factor. By choosing the appropriate momentum, one retains only the expanding saddle point solution. Should we instead choose to integrate over both signs for the lapse, the effect is to allow classical solutions with both signs of the initial momentum.  We show that there is in this case an additional relevant saddle point, representing a bouncing background. Hence, in order to avoid the problematic bouncing saddle point solutions, we are forced to use the Lorentzian propagator, as advocated and explained in our earlier works \cite{Feldbrugge:2017kzv,Feldbrugge:2017mbc}. We can, in this way, recover a description of inflation in terms of QFT in curved spacetime. However, we emphasize that a prior, pre-inflationary phase of the right type must be assumed in order to complete inflation as a consistent framework for cosmology. 

The plan of this paper is as follows: to set the scene, we will briefly review the standard description of perturbations in inflation, using QFT in curved spacetime. Then, in section \ref{sec:path integral}, we introduce the path integral formalism for gravity, both for the background and the perturbations. Here we will show that explicit mode functions for the perturbations can be found for all values of the lapse function, a feature that enables us to carry out novel analytical calculations. In section \ref{sec:vanishing} we explore the consequences of taking a vanishingly small initial three-geometry, and establish our result that this limit inevitably leads to unstable fluctuations with an inverse Gaussian distribution (in linear cosmological perturbation theory -- see also \cite{Hofmann:2019dqu} for related work that supports this conclusion). We verify this result by relating it to the instability of the no-boundary proposal in section \ref{sec:nb}. Faced with this problematic result, we explore possible resolutions in section \ref{InitialConditions}. Introducing a pre-inflationary localized, expanding state, we recover the standard description in terms of a single classical background. We furthermore show, again by assuming an approproate initial quantum state, that stable, Gaussian-distributed perturbations can also be recovered. We discuss our results, their implications and future directions in section \ref{sec:discussion}.

\section{QFT in curved spacetime -- common intuition} \label{sec:QFT}

Let us begin by reviewing the standard inflationary calculation, and the intuition that goes along with it. For technical clarity we  work in de Sitter spacetime, regarded as the ``no-roll'' limit of inflation. However all our results carry over to the case of slow-roll inflation with only minor modifications. The great advantage of considering exact de Sitter spacetime is that we can provide many simple analytic formulae. In the flat slicing and in terms of conformal time $\eta,$ the de Sitter metric is given by $ds^2 = a^2 \left( -d\eta^2 + d{\bf{x}}^2 \right)$ with scale factor $a(\eta)= -\frac{1}{H\eta}$, where the Hubble rate $H$ is related to the cosmological constant $\Lambda$ via $H^2 = \Lambda/3$, and the conformal time $\eta$ ranges over $(-\infty , 0)$. Gauge-invariant perturbations in de Sitter spacetime are described by the Mukhanov-Sasaki equation \cite{Kodama:1985bj,Mukhanov:1988jd}
\begin{equation}
v_k^{\prime\prime} + \left(k^2 - \frac{2}{\eta^2} \right) v_k = 0\,,
\end{equation}
with $v_k$ the canonically normalized perturbation in Fourier space and $k$ the magnitude of the corresponding wave number. These perturbations can be thought of as the time-dependent part of gravitational waves, or as arising from a massless scalar field representing the fluctuations of the inflaton about a slow-roll solution. The equation of motion admits two solutions, one of positive and one of negative frequency
\begin{equation}
v_k = c_1\, e^{-ik\eta}\left(1 - \frac{i}{k\eta} \right) + c_2\, e^{+ik\eta}\left(1 + \frac{i}{k\eta} \right)\,,
\end{equation}
where $c_{1,2}$ are integration constants. In order for quantum field theory in curved spacetime to describe the initial quantum state it is important to select the appropriate mode function. It is usually argued that in the far past, \textit{i.e.} in the limit $\eta \rightarrow - \infty,$ the equation of motion becomes that of a fluctuation in Minkowski spacetime. Consequently, gravity is assumed to become unimportant at early times. Moreover, since the stable, positive frequency solution to the wave equation in Minkowski spacetime is of the form $v_k \propto e^{-ik\eta}$, the condition
\begin{equation}
\lim_{\eta \to -\infty} v_k(\eta) = \sqrt{\frac{\hbar}{2k}} e^{-ik\eta}
\end{equation}
leads one to set $c_2=0$ so that one obtains the Bunch-Davies vacuum \cite{Bunch:1978yq}\footnote{A number of cosmologists have pointed out the dangers of this assumption in the past, see in particular the description of the transPlanckian problem in \cite{Martin:2000xs}.} 
\begin{equation}
v_k = \sqrt{\frac{\hbar}{2k}} e^{-ik\eta}\left(1 - \frac{i}{k\eta} \right) \,.
\end{equation}
The Bunch-Davies vacuum quantum fluctuations are then amplified into a Gaussian distribution of late-time fluctuations $v_k/a$ that reach a constant value  on super-horizon scales and exhibit a scale-invariant spectrum ($|v_k/a|^2 \propto \hbar \,H^2/k^3$). Within the framework of QFT in curved spacetime, this argument seems to indicate that a a universe which starts out in an initial (quasi-) de Sitter inflationary phase nicely matches current observations.

\section{Semi-classical gravity} \label{sec:path integral}

We now wish to reconsider this calculation in semi-classical gravity, meaning that we should evaluate the Feynman path integral amplitude 
\begin{equation}
G[h_{ij}^1,\varphi_1;h_{ij}^0,\varphi_0] = \int \mathcal{D} g\, \mathcal{D}\varphi \ e^{\frac{i}{\hbar} S[g,\varphi]}\label{eq:Tprop}
\end{equation}
to propagate from an initial three-geometry with metric $h_{ij}^{0}$ and, potentially, matter fields $\varphi_0$ to a final three-geometry with metric $h_{ij}^{1}$ and matter fields $\varphi_1$. The action $S$ is taken to be the Einstein-Hilbert action $S_{EH}[g]$ along with a matter action $S_m[\varphi,g]$. The path integral is performed over all four-metrics $g$ and matter fields $\varphi$ consistent with the specified boundary conditions. Note that the expression given is only formal due to the diffeomorphism invariance of the Einstein-Hilbert action, leading to an over-counting of four-geometries: in general a proper treatment requires ghosts. However, for the case at hand, where we shall treat the fluctuations only at quadratic order, the simpler description given here suffices. 

For a homogeneous and isotropic background universe with only small fluctuations $\phi$ (which in the simplest case consist only of gravitational waves), the background spacetime can be described by two zero modes (moduli) namely the lapse $n(t)$ and the scale factor $a(t)$: the line element is given by
\begin{equation}
\mathrm{d}s^2 = -n (t)^2 \mathrm{d}t^2 + a(t)^2 \gamma_{i j} (x) dx^i dx^j,
\end{equation}
where $\gamma_{i j} (x)$ is the three-metric for a maximally symmetric space. In this paper, for the most part we shall consider a spatially flat, FRW cosmology with $\gamma_{i j} (x)=\delta_{ij}$ and a toroidal topology, {\it i.e.}, periodic boundary conditions in comoving coordinates. In Section V we generalize this to a spherical three-geometry for comparison. In these variables, using BFV quantization \cite{Batalin:1977pb} one may impose the proper-time gauge $\dot{n}=0$, and the Feynman propagator \eqref{eq:Tprop}  can be rigorously  expressed as
\begin{equation}
G[a_1,\phi_1;a_0,\phi_0] = \int_{0^+}^\infty \mathrm{d}n 
\int_{a_0}^{a_1} \mathcal{D} a\
\int_{\phi_0}^{\phi_1}
\mathcal{D} \phi\ e^{\frac{i}{\hbar}S[g,\phi;n]}
\end{equation} 
as derived by Teitelboim \cite{Teitelboim:1981ua, Teitelboim:1983fk} and Halliwell \cite{Halliwell:1988wc}. Note that in this gauge, the integral over the gauge-fixed lapse $n$ amounts to an integral over the proper time between the initial and the final three-geometry. In this section we first focus on the propagator for the background. In subsequent sections, we analyze the  fluctuations. 

\subsection{The background}\label{sec:background}

To set the scene, recall the description of the classical background -- de Sitter spacetime in the flat slicing. Taking the line element as $\mathrm{d}s^2 = -n(t)^2 \mathrm{d}t^2+a(t)^2 \mathrm{d}{\bf x}^2$, the action is
\begin{equation}
{S}^{(0)}= V_3 \int_0^1 dt \left(-3 M_{Pl}^2 n^{-1} a \dot{a}^2  -n a^3 \Lambda\right),
\label{frwact}
\end{equation}
where $M_{Pl}^2\equiv 1/(8 \pi G)$ is the reduced Planck mass and $\Lambda$ is the cosmological constant which we shall assume to be positive. To avoid clutter, we generally set $M_{Pl}=1$ in what follows, restoring it by dimensions when helpful. Without loss of generality we may choose the coordinate $t$ to run from $0$ to $1$ and the spatial volume  $V_3$ of the torus, in comoving coordinates, to be unity. The action (\ref{frwact}) is inconvenient because it is cubic in $a$ and $\dot{a}$. However, if we redefine $N(t)=a(t) n(t)$ and $q(t)=a(t)^2$, so that the line element is $\mathrm{d}s^2 = -N(t)^2 \mathrm{d}t^2/q(t)+q(t) \mathrm{d}{\bf x}^2$, the action becomes quadratic~\cite{Halliwell:1988ik} and hence easier to analyze\footnote{As we are considering semi-classical gravity, we will ignore factor ordering ambiguities and Jacobian factors in the measure, which will lead to corrections at subleading orders in $\hbar$.} :
\begin{equation}
{ S}^{(0)}= \int_0^1 dt \left(-{3\over 4}  N^{-1} \dot{q}^2  - N q \Lambda\right),
\label{frwact2}
\end{equation}
showing that $q$ behaves as the coordinate of a particle moving in a linear potential. The canonical momentum conjugate to $q$ is $p=-{3\over 2} N^{-1} \dot{q} $.  In natural units, the  coordinates $t$ and ${\bf x}$ are dimensionless, $n$ and $a$ have dimensions of length, $N$ and $q$ have dimensions of length$^2$ and $p$ has dimensions of length$^{-2}$. These scalings are helpful in our later analysis in section \ref{sec:stableperts}. The Hamiltonian is $H=N\left(-{1\over 3} p^2 +q \Lambda\right)$. It vanishes, when the equations of motion are satisfied, as a consequence of time reparameterization invariance. 

\begin{figure}[h]
\centering
\begin{minipage}{0.45\textwidth}
\includegraphics[width=\linewidth]{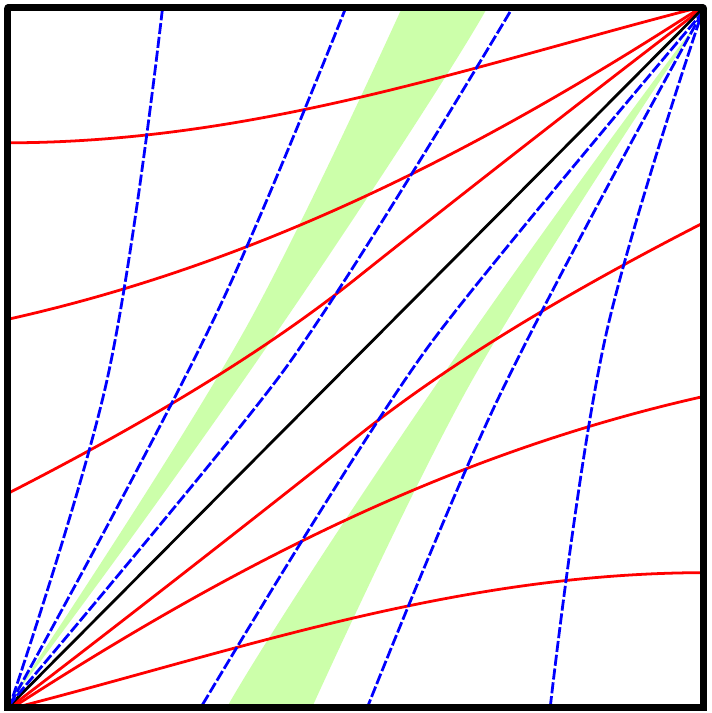}
\end{minipage}~
\begin{minipage}{0.45\textwidth}
\includegraphics[width=\linewidth]{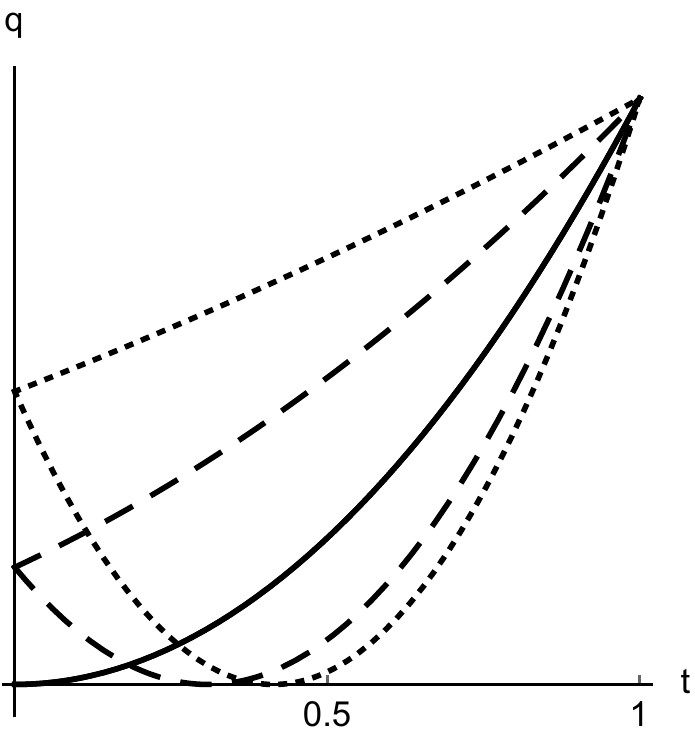}
\end{minipage}~
\caption{{\it Left panel:} The Penrose diagram of de Sitter space-time in the flat slicing. The red lines denote time slices and the blue lines denote space slices. The two background solutions relevant to the propagator consist of either pure expansion (a finite portion of the green patch on the left) or first contraction followed by expansion (a portion of the green patch on the right disappearing at the top right, reappearing bottom left and ending up at the same location as the purely expanding solution). {\it Right panel:} The two background solutions relevant to the propagator, shown as a function of $t$ for successively decreasing initial scale factor squared $q_0$ (dotted, then dashed, and finally solid at $q_0=0$). As the initial scale factor is decreased, the two solutions approach each other and merge when $q_0=0.$ This last solution would correspond to the limit where one starts at the bottom left corner (or equivalently the top right corner) in the Penrose diagram on the left.}\label{fig:bgd}
\end{figure}

Evidently $q$, the scale factor squared, behaves like the coordinate of a particle moving with zero energy in a linear potential. It will be convenient to pick a gauge in which the lapse is constant, $\dot{N}=0$. Classically, the value of $N$ encodes the total proper time between the initial and final three-geometry. Quantum mechanically, when performing the path integral we must integrate over all positive values of $N$. Clearly, for a linear potential $-q \Lambda$ there are two classical solutions which travel, at zero energy, from some initial $q_0$ to a final, larger $q_1$. Either $q$ increases all the way or it starts out decreasing, ``bounces" off the potential at $q=0$, and increases to $q_1$. These are the two  trajectories, alluded to in the introduction, which we shall study semi-classically in some detail, along with their associated perturbations. 

The existence of the ``bouncing" trajectory, with a larger real value of $N$ and, correspondingly, a larger proper time,  is closely related to the fact that the flat slicing only covers half of de Sitter spacetime. In a bouncing trajectory, $q$ vanishes as $t^2$ near the bounce so the line element behaves (up to constants) as $-\mathrm{d}t^2/t^2+t^2 \mathrm{d}{\bf x}^2$. Defining the conformal time $\eta=-1/t$, the line element becomes $\eta^{-2}(-\mathrm{d}\eta^2 +\mathrm{d}{\bf x}^2)$, the familiar expression for de Sitter spacetime in the flat slicing. As $t$ runs from $-\infty$ to $+\infty$ through all real values, we have an infinite universe collapsing to zero size and rebounding to infinity. This includes both the contracting and expanding halves of de Sitter: see Fig.~\ref{fig:bgd}. We see this by analytically continuing in $t$ (or $\eta$) around the point $t=0$. The given range of $t$ corresponds to $\eta$ running from $0^-$ to $-\infty$, through (or around) the point at infinity and from  $+\infty$ to $0^+$. Standard treatments of inflation are able to ignore one half of de Sitter spacetime by treating the background as classical. However, as we shall show,  when the background is treated quantum mechanically, the bouncing solutions are in general relevant and must be included.

We shall show later that the ``bouncing" trajectory leads to a disastrous probability distribution for perturbations -- either gravitational waves or scalar density perturbations -- and hence must somehow be made irrelevant through a choice of initial conditions. Our main result, that there is a minimum initial size for an inflating patch, follows from this consideration. 
Consider a quantum mechanical particle with coordinate $q$ moving in a potential $-\Lambda q$,  with a positive initial velocity. We can describe it with a wavepacket, assumed Gaussian for convenience, with central value $q_i$, standard deviation $ \sigma\equiv \sqrt{\langle \Delta q^2\rangle}$ and central value of the momentum $p_i$. In order that this initial quantum wavefunction describes a real, classical inflating universe with high probability, we must have $q_i \gg \sigma$ so that the spacetime metric has the desired signature (recall $q\equiv a^2$), we must take $p_i$ to be the expanding solution of the Hamiltonian constraint (Friedmann equation), $p_i=-\sqrt{3 q_i \Lambda}$, and also impose that the uncertainty $\Delta p \sim \hbar/\sigma$ given by the Heisenberg uncertainty relation, is smaller than $|p_i|$ in order that we can be sure that the initial universe is really expanding. Writing $q_i =\nu \sigma$ and $|p_i|=\nu \Delta p$, with $\nu\gg1$ measuring the number of standard deviations by which we can be assured that the initial $q_i$ is positive and that the initial universe is expanding, neglecting numerical factors we find $\sqrt{q_i \Lambda}>\nu^2 \hbar /q_i$. Rearranging this, we find the initial comoving volume of our background torus has to satisfy
 \begin{equation}
V_3 a_i^3 > {\nu^2 \hbar \over M_{Pl} \sqrt{\Lambda}},
\label{estbd}
\end{equation}
where we restored the coordinate three volume and the Planck mass. We shall rederive this bound much more carefully and rigorously in section VI, exhibiting a Stokes phenomenon whereby the ``bouncing" solution becomes irrelevant as the initial size of the universe is raised. These more detailed considerations confirm the scaling exhibited in (\ref{estbd}) and, furthermore, yield an accurate numerical coefficient. 

How large should we take $\nu$ to be? Recall that {\it any} admixture of the ``bouncing" background results in disastrous perturbations. Therefore, to be conservative, such a background should be excluded for {\it all} perturbation modes we consider, that is, all the modes which exited the Hubble radius during inflation, and are now encompassed by region we currently observe. The number of such modes is proportional to $e^{3 N_I}$ where $N_I$ is the number of efoldings of inflation our current Hubble volume underwent.
If we divide this volume up into $e^{3 N_I}$ identical cubes, we need to ensure that none of them underwent the ``bouncing" evolution. The probability for any one cubical region to ``bounce" is suppressed by  $e^{-\nu^2/2}$ for our assumed Gaussian distribution. Therefore, in order to get {\it no} bouncing region we require $\nu> \sqrt{6 N_I} \sim 20$ for $N_I=60$ efolds of inflation. For an inflationary scale $\Lambda^{1\over 4}$ of order the GUT scale $\sim 10^{-3} M_{Pl}$, Eq. (\ref{estbd}) requires an initial inflating volume of around $10^8$ Planck volumes.  

Finally, let us note that the estimate given above may well be generous to inflation. What we have done is treat the isotropic moduli,  {\it i.e.}, the scale factor and the lapse, non-perturbatively, but all other modes perturbatively. Our analysis could, with some effort, be extended to treat the other homogeneous modes -- the {\it anisotropy} moduli -- non-perturbatively as well. Since the inclusion of anisotropies tends to counteract inflation and strengthen the onset of singularities, this may make it harder to choose quantum initial conditions which avoid singular semiclassical trajectories of the type we have shown to lead to uncontrolled perturbations.

\subsection{Background path integral and saddles}\label{sec:backgroundpi}

The Feynman propagator for the background, in these variables, is
\begin{align}
G[q_1;q_0] = \int_{0^+}^\infty dN \int_{q_0}^{q_1} {\mathcal D}q\ e^{\frac{i}{\hbar}S^{(0)}[q;N]}\,,
\label{feynprop}
\end{align}
with $S^{(0)}[q;N] = \int_0^1 dt  \left( -\frac{3}{4N} \dot{q}^2 -Nq\Lambda \right)\,$. Since the action is quadratic in $q$, the path integral over the scale factor $q$ can be expressed in terms of the classical action. With the specified boundary conditions, the solutions to the equation of motion $\ddot{q}=\frac{2\Lambda}{3}N^2$ are given by
\begin{equation}
\bar{q}(t) = H^2 N^2 (t - \alpha)(t -\beta),
\end{equation}
with
\begin{equation} 
\alpha,\beta = \frac{1}{2 N^2} \left[N^2 - N_{s-}N_{s+} \pm \sqrt{( N^2 - N_{s-}^2)( N^2 - N_{s+}^2 )} \right] \,, \label{ab}
\end{equation}
where
\begin{equation}
N_{s\pm} = \frac{\sqrt{q_1} \pm \sqrt{q_0}}{H}\,.\label{eq:saddle}
\end{equation}
The corresponding classical action is given by
\begin{align}
\bar{S}^{(0)}[q_1;q_0;N] = V_3 \left[ \frac{\Lambda^2}{36}N^3 -\frac{\Lambda}{2}(q_0+q_1)N -\frac{3(q_1-q_0)^2}{4N}\right]\,,
\end{align}
where $V_3$ denotes the spatial three-volume at $q=1,$ which we assume to be finite. In the subsequent calculation we will choose spatial coordinates such that $V_3=1,$ though we will re-instate $V_3$ explicitly in section \ref{InitialConditions}. Using the classical action $\bar{S}^{(0)}$, the Feynman propagator reduces to an oscillatory integral over the lapse in the proper-time gauge
\begin{align}
G[q_1;q_0] = \sqrt{\frac{3 i}{4\pi \hbar}} \int_{0^+}^\infty \frac{\mathrm{d}N}{\sqrt{N}}  e^{\frac{i}{\hbar}\bar{S}^{(0)}[q_1;q_0;N]}\,. \label{Nintegral}
\end{align}

\begin{figure}
\begin{minipage}{0.5\textwidth}
\includegraphics[width=\linewidth]{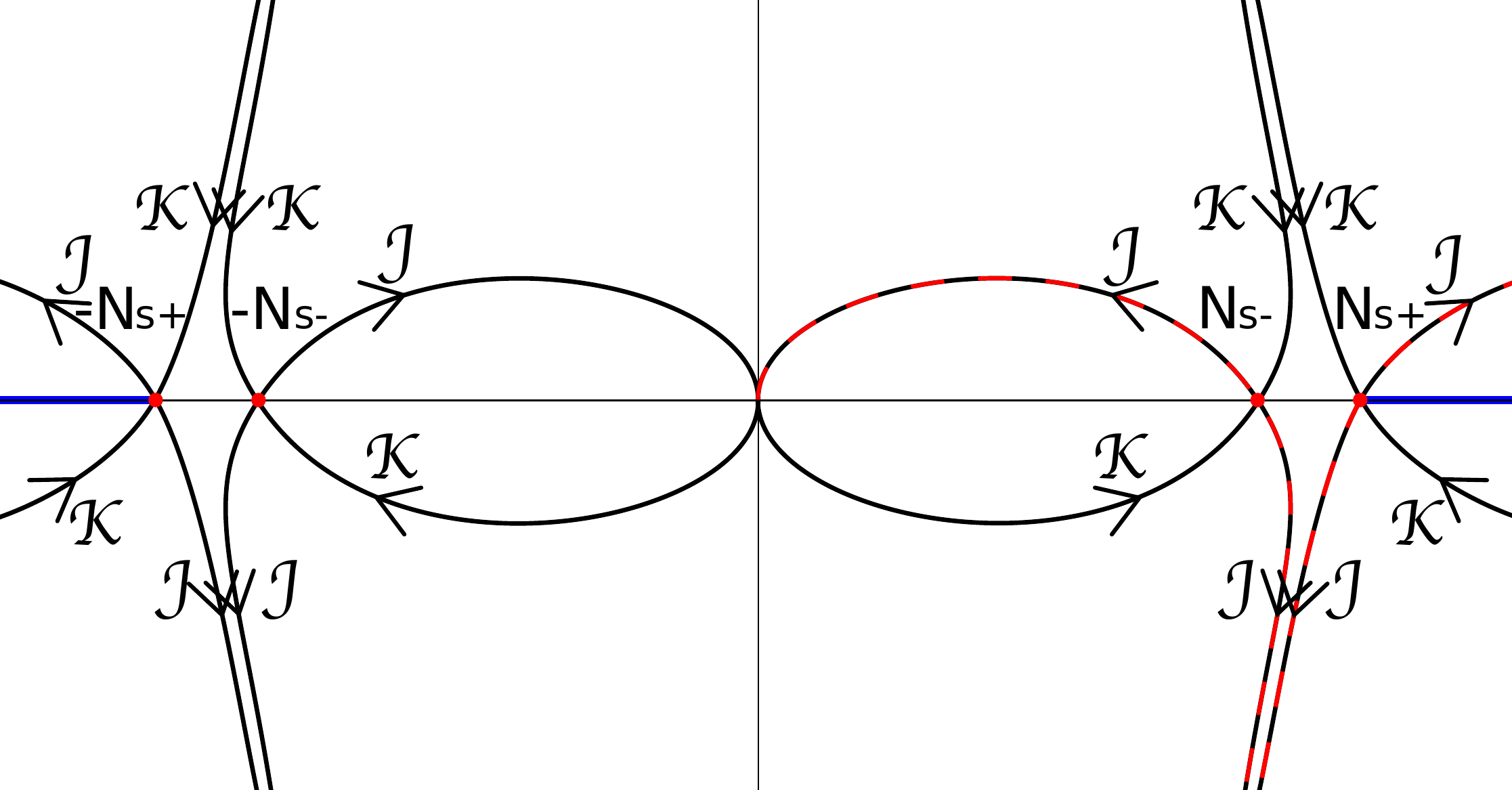}
\end{minipage}~
\begin{minipage}{0.5\textwidth}
\includegraphics[width=\linewidth]{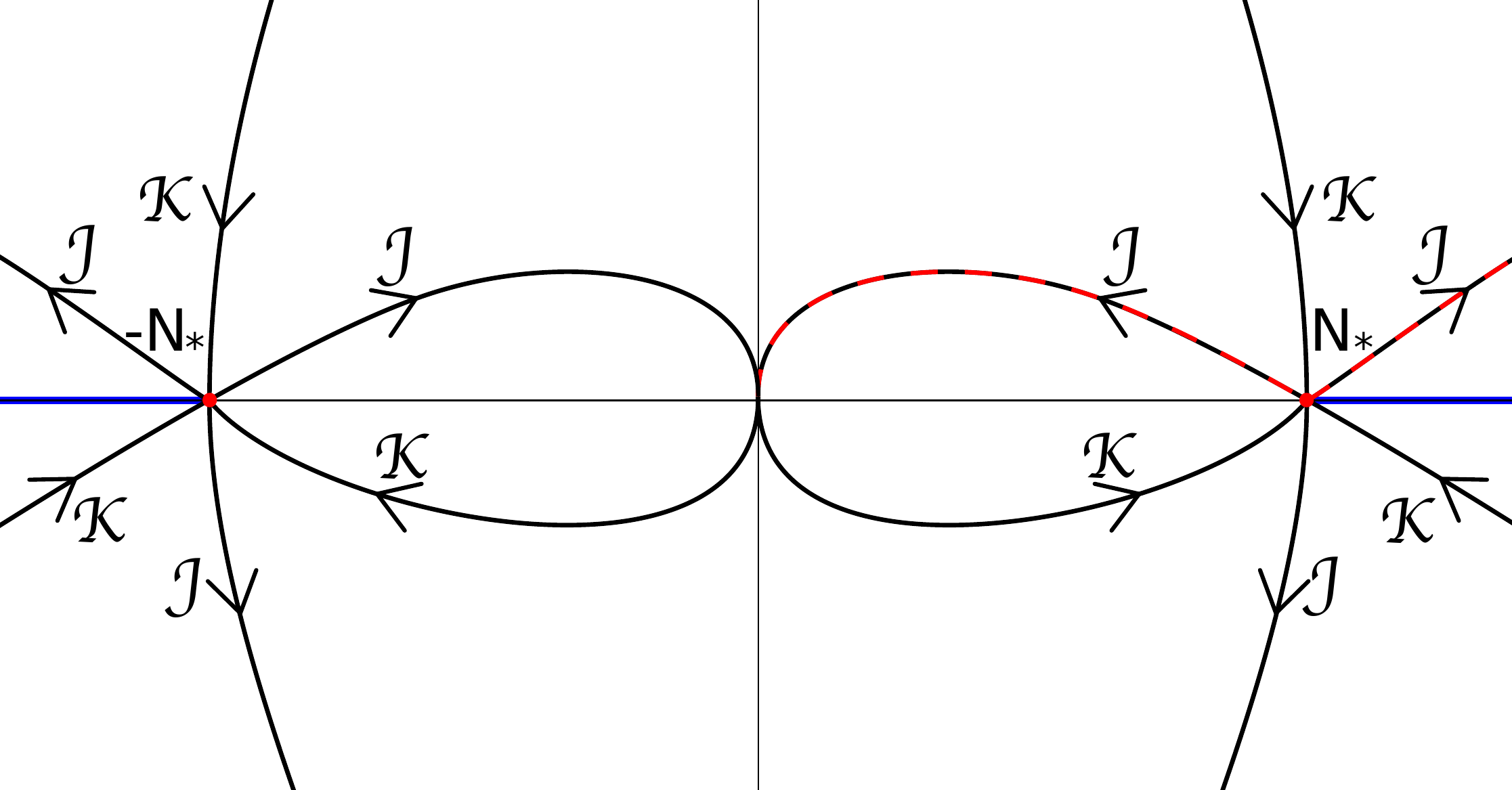}
\end{minipage}
\caption{The lines of steepest ascent $\mathcal{K}$ and descent $\mathcal{J}$ for the lapse integral, shown in the complexified plane of the lapse $N$. Left: for generic boundary conditions with $q_0,q_1 \neq 0$. Right: the limit of vanishing initial boundary $q_0=0$.}\label{fig:PLbackground}
\end{figure}

We approximate the lapse integral in the saddle point approximation using Picard-Lefschetz theory \cite{Feldbrugge:2017kzv}.
The exponent $i \bar{S}^{(0)}/\hbar$ has four saddle points in the lapse $N$, located at $\pm N_{s\pm}$. The lines of steepest ascent and descent emanating from the saddle points run through the complex plane to the essential singularities at the origin and complex infinity (see figure \ref{fig:PLbackground}). Since we are integrating the lapse over the positive real line, only the two saddle points with positive real part $+N_{s\pm}$ are relevant to the integral (assuming $q_0 \leq q_1$ as we will henceforth). For generic boundary conditions, with $q_0,q_1 \neq 0$, the saddle points are non-degenerate. The saddle point approximation of the Feynman propagator is given by
\begin{equation} 
G[q_1;q_0] \approx \sqrt{ \frac{3i}{4\Lambda \sqrt{q_0 q_1}}}
 \left[ \, e^{-i\frac{\pi}{4}} e^{i\bar{S}^{(0)}[N_{s-}]/\hbar} +   \, e^{i\frac{\pi}{4}}e^{i\bar{S}^{(0)}[N_{s+}]/\hbar} \right]\,,
\end{equation}
with the classical action at the saddle points
\begin{equation}
\bar{S}^{(0)}[N_{s\pm}]=   -2 \sqrt{\frac{\Lambda}{3}}  \left( q_1^{3/2} \pm q_0^{3/2} \right)\,.
\label{nbwf_classical}
\end{equation} 

The propagator consists of the interference of two classical solutions. At the relevant saddle points  $N_{s\pm}$, the background solutions simplify to 
\begin{align}
\bar{q}\mid_{N_{s\pm}} = H^2 N_{s\pm}^2 (t-\alpha)^2\,, \qquad \alpha = \beta = \pm \frac{\sqrt{q_0}}{HN_{s\pm}}\,.
\end{align} 
At $N_{s-}$, the parameter $\alpha$ is negative and the background continuously expands from $q_0$ to $q_1$. In physical time $t_p$ this expansion takes the usual exponential form $a(t_p)=\frac{1}{H}e^{+Ht_p}$ for a suitable range of $t_p$. At $N_{s+}$ the parameter $\alpha$ is positive and the universe first contracts from $q_0$ to zero size and then re-expands to $q_1$. In proper time $t_p$ this amounts to a contracting phase given by $a(t_p) = \frac{1}{H} e^{-H t_p}$ followed by an expanding phase described by $a(t_p) = \frac{1}{H} e^{+H t_p}$ (see figure \ref{fig:bgd}). In QFT on curved spacetime one only works with the solution corresponding to the inner saddle point $N_{s-}$, restricting the calculation to the upper triangle of the Penrose diagram of de Sitter spacetime in the flat slicing (see the upper left triangle of figure \ref{fig:bgd}). In quantum gravity we cannot restrict our analysis to a fixed evolution of the scale factor of de Sitter space time and need to consider the solutions corresponding to both $N_{s-}$ and $N_{s+}$ located in the entirety of de Sitter spacetime.

As we saw in section \ref{sec:QFT}, the Bunch-Davies vacuum is selected by considering the limit $\eta \to -\infty$. In semi-classical gravity, this corresponds to the limit where the initial boundary shrinks to a point $q_0 \rightarrow 0$,
while $q_1$  is held fixed at a large positive value. This is an interesting limit for the background universe since the two classical solutions coincide and the two saddle points $N_{s\pm}$ merge, thus forming a degenerate saddle point of order $2$ located at
\begin{align}
N_\star \equiv \sqrt{\frac{3q_1}{\Lambda}} = \frac{\sqrt{q_1}}{H}\,\,.
\end{align}
See figures \ref{fig:bgd} and \ref{fig:PLbackground} for an illustration. The saddle point approximation to the Feynman propagator in this limit involves an integral over cubic fluctuations around the (degenerate) saddle point, as explained in~\cite{Feldbrugge:2017kzv}, and reads
\begin{equation} \label{nbwf_classical_limit}
G[q_1;q_0=0]  \approx \frac{e^{i\frac{\pi}{4}}3^{17/12}\Gamma(\frac{4}{3})}{2 \pi^{1/2}\hbar^{1/6}\Lambda^{5/12}q_1^{1/4}} \, e^{-i2\sqrt{\frac{\Lambda}{3}} q_1^{3/2}/\hbar} \,. 
\end{equation} 
We thus observe that generic boundary conditions lead to an interference of an expanding and a bouncing solution. In the limit of a vanishing initial boundary, however, the interference changes form. The propagator is now dominated by a single classical solution corresponding to a degenerate saddle point in the saddle point approximation. 

For completeness, let us note that the path integral \eqref{Nintegral}) admits an exact expression in terms of products of Airy functions. It can be shown (see for instance \cite{Halliwell:1988ik,Feldbrugge:2017kzv}) that it satisfies the inhomogeneous WdW equation 
\begin{equation}
\hbar^2 \frac{\partial^2 G[q_1 ; q_0]}{\partial q_1^2} + 3 (\Lambda q_1 - 3 k ) G[q_1 ; q_0] = - 3 i \hbar \,  \delta(q_1 - q_0)\,.
\end{equation}
Defining $z \equiv \frac{(-3)^{1/3}}{(\hbar \Lambda)^{2/3}} (\Lambda q - 3 k),$ this equation is solved by 
\begin{equation}
G[q_1 ; q_0] = \frac{\pi (-3)^{2/3}}{(\hbar \Lambda)^{1/3}} \{ Ai(z_1) [Ai (z_0) - i Bi(z_0) ] \Theta(q_1 - q_0) + Ai(z_0) [Ai(z_1 )- i Bi(z_1)] \Theta(q_0 - q_1) \}
\end{equation}
The above expression then reduces to (\ref{nbwf_classical_limit}) for $k=0$, $q_0 = 0$ in the limit of large $q_1$.

It is important to note that the degeneracy of the saddle point in $N$ in the limit $q_0 \to 0$ stems from the fact that we do not a priori know whether the boundary conditions of the propagator lie in the upper or lower triangle of the Penrose diagram (see figure \ref{fig:bgd}). Since the analytically continued scale factor $a$ changes sign at the singularity, one might argue that the degeneracy can be lifted by specifying the relative sign of the scale factors $a_0$ and $a_1$. However it should be noted that the the metric -- over which we integrate in the path integral -- is insensitive to the sign of the scale factor. It is for this reason appropriate to work in terms of the squared scale factor $q=a^2$. Moreover, in section \ref{sec:nb} we derive the same result from the no-boundary proposal in the limit of vanishing curvature, where no such ambiguities exist as the scale factor is everywhere non-negative.

\subsection{The fluctuations} \label{subsec:fluctuations}

To leading order, the perturbations are described by the action
\begin{align}
S^{(2)}[\phi,q;N] = \frac{1}{2} \int_0^1 dt \int d^3x \left( \frac{q^2}{N}\dot\phi_k^2 - N k^2 \phi_k^2 \right)\,,
\end{align}
where we focus on a single mode with wavenumber $k$. A sum over Fourier modes is straightforward to implement, but will be kept implicit. One may think of the fluctuation as the component of a gravitational wave, or an additional massless scalar field. The equation of motion, ignoring the backreaction of the fluctuations on the metric, is given by
\begin{equation}
\ddot{\phi} + 2 \frac{\dot{\bar{q}}}{\bar{q}} \dot{\phi}+ \frac{k^2 N^2}{\bar{q}^2} \phi = 0 \label{212}\,.
\end{equation}
The solutions are of the form
\begin{equation}
\phi(t) = \frac{a f(t) + b g(t)}{\sqrt{\bar{q}(t)}} \label{linearcomb}
\end{equation}
with
\begin{equation}
f(t) = \Bigl[\frac{t - \beta}{t - \alpha}\Bigr]^{\mu/2}   \Bigl[(1 - \mu)(\alpha - \beta) + 2(t - \alpha) \Bigr]\,,
\end{equation}
\begin{equation}
g(t) = \Bigl[\frac{t - \alpha}{t - \beta }\Bigr]^{\mu/2}   \Bigl[(1 + \mu)(\alpha - \beta) + 2(t - \alpha)\Bigr]\,.
\end{equation}
Here the exponent $\mu$ is given by
\begin{equation}
\begin{split}
\mu^2 &= 1 -\Bigl( \frac{2k  }{ (\alpha - \beta)  H^2 N}\Bigr)^2 \\
& \equiv \frac{(N^2 - \overline{N}_{-}^2)(N^2 - \overline{N}_{+}^2)}{(N^2 - N_{s-}^2)(N^2 - N_{s+ }^2)}\,, \label{defmu}
\end{split}
\end{equation}
where the zeros of $\mu$ are specified by
\begin{equation}
\pm \overline{N}_{\pm} = \pm \frac{\sqrt{q_1^2 H^2 + k^2} \pm \sqrt{q_0^2 H^2 + k^2}}{H^2} 
\end{equation}
The integration constants $a$ and $b$ in the solution \eqref{linearcomb} can be determined by imposing the boundary conditions $\phi(t=0) = \phi_0$ and $\phi(t=1)=\phi_1,$ leading to
\begin{equation}
a =+  \frac{1}{D}\left[ \left[\frac{1 - \alpha}{1 - \beta }\right]^{\mu/2} (2 - (\alpha + \beta) + \mu (\alpha - \beta)) \sqrt{q_0} \phi_0 + \left[\frac{ \alpha}{\beta }\right]^{\mu/2} (\alpha + \beta - \mu (\alpha - \beta)) \sqrt{q_1} \phi_1 \right]
\end{equation}
\begin{equation}
b = - \frac{1}{D} \left[ \left[\frac{1 - \beta}{1 - \alpha}\right]^{\mu/2} (2 - (\alpha + \beta) - \mu(\alpha - \beta))\sqrt{q_0} \phi_0 + \left[\frac{\beta}{ \alpha}\right]^{\mu/2} (\alpha + \beta + \mu (\alpha  - \beta)) \sqrt{q_1} \phi_1 \right]
\end{equation}
\begin{equation}
\begin{split}
D =&+ \left[\frac{(1 - \beta)\alpha}{(1 - \alpha) \beta}\right]^{\mu/2} (2 - (\alpha + \beta) - \mu(\alpha - \beta))(\alpha + \beta - \mu(\alpha - \beta))  \\
&- \left[\frac{(1 - \alpha)\beta}{(1 - \beta) \alpha }\right]^{\mu/2}  (2 - (\alpha + \beta) + \mu(\alpha - \beta))(\alpha + \beta + \mu(\alpha - \beta))\,.
\end{split}
\end{equation}

\begin{figure}
\includegraphics[width=0.4 \textwidth]{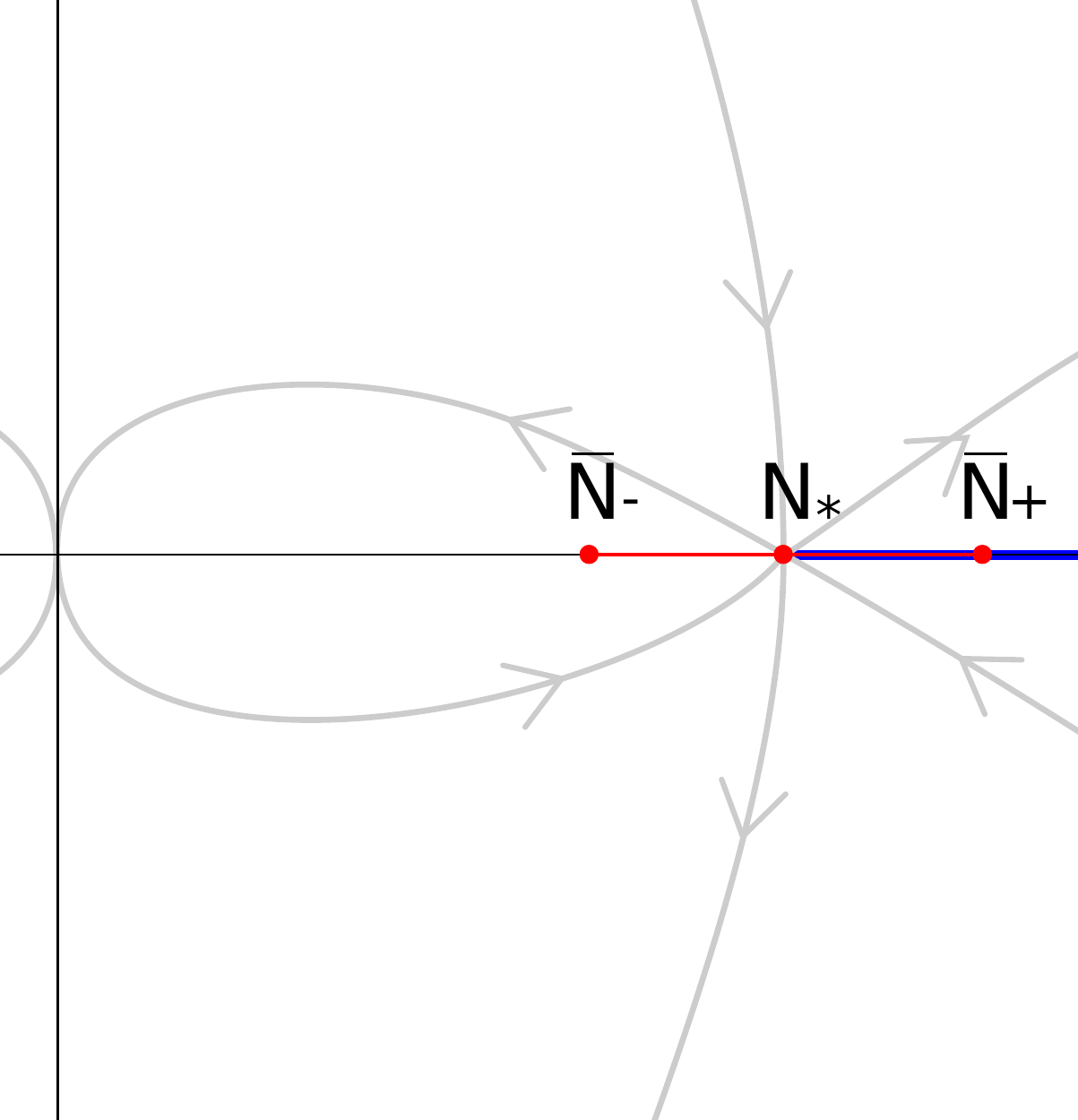}
\caption{The complex $N$ plane with the degenerate saddle point $N_*$ and the branch points $\bar{N}_{+}$ and $\bar{N}_{-}$, relevant to the case where the initial size of the universe is taken to zero. The red line represents the branch cut. The blue line represents the $N$ for which the background $\bar{q}$ passes through $0$ for a time $t \in (0,1]$.}\label{fig:plot}
\end{figure}

Since the perturbation action $S^{(2)}$ is quadratic in $\phi$, the classical action $\bar{S}^{(2)}$ is given by the boundary terms
\begin{equation}
\bar{S}^{(2)}[\phi_1,q_1;\phi_0,q_0] = \frac{\bar{q}^2(t) \phi(t) \dot{\phi}(t)}{2 N} \bigg|_{t=0}^1\,. \label{eq:boundary}
\end{equation}
Equation \eqref{eq:boundary} holds for all $N$ in the complex plane except for part of the real line $|N| \geq N_{\star}$, where $q$ passes through zero and additional singularities appear (see figure \ref{fig:plot} and a closely related discussion in \cite{Feldbrugge:2017mbc}). These parts of the real $N$ line must be excluded from the domain of integration in the integral, since the action becomes infinite there. Away from these line segments, the action is explicitly given by
\begin{equation}
\bar{S}^{(2)} = \frac{n}{d}
\end{equation}
with the numerator
\begin{align*}
n = \frac{ k^2}{H^2 N } \Bigl \{ &+ \left [\frac{(1 - \alpha) \beta}{(1 - \beta) \alpha} \right]^{\mu/2} \left[ (2 - (\alpha + \beta) + \mu (\alpha - \beta))q_0 \phi_0^2  + (\alpha + \beta + \mu (\alpha - \beta))q_1 \phi_1^2\right]   \\
& - \left [\frac{(1 - \beta) \alpha}{(1 - \alpha) \beta} \right]^{\mu/2} [(2 - (\alpha + \beta) - \mu (\alpha - \beta)) q_0 \phi_0^2  + (\alpha + \beta - \mu (\alpha - \beta)) q_1 \phi_1^2  ]  \\
& - 4 \mu (\alpha - \beta) \sqrt{q_1 q_0}\phi_1 \phi_0 \Bigr \}\,,
\end{align*}
and the denominator
\begin{align*}
d =& +   \left [\frac{(1 - \beta) \alpha}{(1 - \alpha) \beta} \right]^{\mu/2} ( 2 -( \alpha + \beta) - \mu (\alpha - \beta)  )( \alpha + \beta - \mu (\alpha - \beta )) \\
& -  \left [\frac{(1 - \alpha) \beta}{(1 - \beta) \alpha} \right]^{\mu/2} (+ 2 -( \alpha + \beta) + \mu (\alpha - \beta)  )( \alpha + \beta + \mu (\alpha - \beta ))\,.
\end{align*}

Despite the occurrence of roots in $\alpha, \beta$ and $\mu,$ the action does not contain branch points as long as $q_0, q_1 \neq 0.$ To see this, it is useful to express the root as an explicit function of $N$,
\begin{equation}
\left [\frac{(1 - \alpha) \beta}{(1 - \beta) \alpha} \right]^{\mu/2} =  \left [\frac{N_{s+}^2 + N_{s-}^2 - 2 N^2 + 2 \sqrt{(N^2 - N_{s+}^2)(N^2 - N_{s-}^2)}}{N_{s+}^2 + N_{s-}^2 - 2 N^2 - 2 \sqrt{(N^2 - N_{s+}^2)(N^2 - N_{s-}^2)}} \right]^{ +\frac{1}{2} \sqrt{\frac{(N^2 - \overline{N}_{+}^2)(N^2 - \overline{N}_{-}^2)}{(N^2 - N_{s+}^2) (N^2 - N_{s-}^2)}}} \label{root}
\end{equation}
As one considers a closed loop in $N$ which includes for example $N_{s-}$, the numerator and the denominator in this last expression exchange their roles. However, since the exponent changes its sign at the same time, the action remains unchanged, and the saddle points $\pm N_{s\pm}$ are not branch points. Similarly, the net effect of completing a loop around $ \pm \overline{N}_{\pm}$ (\textit{i.e.} changing the sign of $\mu$) is to send  $n \rightarrow - n$ and $d \rightarrow - d$ and once again the action remains unaffected.

\section{The limit of a vanishing initial three-geometry} \label{sec:vanishing}

In section \ref{sec:QFT} we showed that the Bunch-Davies vacuum is selected by assuming that the (quasi-) de Sitter phase extends back to the ``beginning'' of the universe and considering the limit of early conformal time $\eta \to -\infty$. In the path integral formulation it is more natural to consider the limit where the initial scale factor goes to zero $q_0\to 0$. From the explicit mode functions \eqref{linearcomb} it follows that the combination $q_0 \phi_0^2$ appearing in the action tends to zero in this limit. The classical action for the fluctuations simplifies to
\begin{equation}
\bar{S}^{(2)}[\phi_1,q_1;\phi_0,q_0=0;N] = - \frac{k^2 N q_1 \phi_1^2}{H^2} \frac{1}{N^2 + N_\star^2 - \sqrt{(N^2 - \overline{N}_{+}^2)(N^2 - \overline{N}_{-}^2)}}\,.\label{actionzeroq0}
\end{equation}	
Two important consequences emerge:
\begin{itemize}
\item The action now contains branch points at $\pm \overline{N}_{\pm}$. It is convenient to place the associated branch cut on the real $N$ line between $\overline{N}_-$ and $\overline{N}_+$ and similarly on the negative $N$ axis (see figure \ref{fig:plot}). This cut appears only in the limit $q_0 \rightarrow 0$ since the terms that would be necessary to compensate for the sign changes when considering closed loops around $\pm \overline{N}_{\pm}$ are now absent. It turns out that we are interested in the action $\bar{S}^{(2)}$ evaluated at $N_\star$ which lies on the branch cut. In the analysis below we study the action at $N_\star$ in more detail. 
\item The classical action is independent of $\phi_0$. Hence we obtain a unique result for the final fluctuations, regardless of what the initial fluctuations are. In other words, in the limit where the de Sitter phase, or inflation, started the universe, we obtain a unique result for the quantum vacuum. Conversely, we may use this calculation to test whether inflation can be considered as a theory of the initial phase of the universe.
\end{itemize}

In the analysis of the background, we saw that the two saddle points $N_{s\pm}$ merge into a degenerate saddle point at $N_\star$. The fact that the action \eqref{actionzeroq0} contains a branch cut indicates that it may be delicate to evaluate it at $N_*$. It is indeed clearer to start with the case where $q_0 \neq 0$ and only then take the limit where the initial scale factor tends to zero. Moreover, this limiting procedure allows one to make contact with the calculation of inflationary fluctuations in the framework of QFT in curved spacetime. 

The evaluation of the action and the classical solutions at the saddle points must be taken with care, since at the saddle points we have $\alpha = \beta$ while the action contains negative powers of $(\alpha - \beta)$.  Using the limit $\lim_{x\rightarrow \infty}(1+\frac{1}{x})^x = e$, we obtain the following identities
\begin{align}
\mu(\alpha - \beta) & \rightarrow \,  \frac{2ik}{H^2 N_{s\pm}} \\
\left[ \frac{t-\beta}{t-\alpha} \right]^{\mu/2} = \left[1+ \frac{\alpha-\beta}{t-\alpha} \right]^{\mu/2} & \rightarrow \, e^{ \frac{ik}{H^2 N_{s\pm}(t-\alpha)}} \\
\left[ \frac{(1-\beta)\alpha}{(1-\alpha)\beta} \right]^{\mu/2} = \left[1+ \frac{\alpha-\beta}{(1-\alpha)\beta} \right]^{\mu/2} & \rightarrow \, e^{\frac{ik}{H}(\frac{1}{\sqrt{q_1}}\pm \frac{1}{\sqrt{q_0}})} \\
\left[ \frac{\alpha}{\beta} \right]^{\mu/2}= \left[1- \frac{\alpha-\beta}{\beta} \right]^{\mu/2} & \rightarrow \, e^{\pm \frac{ik}{H\sqrt{q_0}}}
\end{align}
at the saddle points $N_{s\pm}$ (where the upper/lower signs are always correlated). The mode functions may be re-expressed as
\begin{align}
\frac{HN_{s\pm}}{2}\frac{f(t)}{\sqrt{q}}\bigg|_{N_{s\pm}} & \rightarrow \,e^{\frac{ik}{H^2 N_{s\pm}(t-\alpha)}} \left(1 - \frac{ik}{H^2 N_{s\pm}(t-\alpha)} \right) \\
\frac{HN_{s\pm}}{2}\frac{g(t)}{\sqrt{q}}\bigg|_{N_{s\pm}} & \rightarrow \,e^{-\frac{ik}{H^2 N_{s\pm}(t-\alpha)}} \left(1 + \frac{ik}{H^2 N_{s\pm}(t-\alpha)} \right)\,. 
\end{align}
These are the familiar mode functions since identifying conformal time $\eta$ as 
\begin{equation}
\eta = -\frac{1}{H\sqrt{\bar{q}}} = - \frac{1}{H^2 N_{s-}(t-\alpha)}
\end{equation}
with $\eta_0 = -\frac{1}{H\sqrt{q_0}}$ and $\eta_1 = -\frac{1}{H\sqrt{q_1}}$\footnote{The boundary conditions in conformal time $\eta_0$ and $\eta_1$ correspond to the expanding solution and should be considered as a short hand for the condition in terms of $q_0$ and $q_1$.}, we recover the standard Bunch-Davies mode functions in conformal time,
\begin{equation}
\frac{HN_{s-}}{2}\frac{f(t)}{\sqrt{\bar{q}}}\bigg|_{N_{s-}}  \rightarrow \,e^{-ik\eta} \left(1 + ik\eta \right)\,,
\quad
\frac{HN_{s-}}{2}\frac{g(t)}{\sqrt{\bar{q}}}\bigg|_{N_{s-}}  \rightarrow \,e^{ik\eta} \left(1 - ik\eta \right)\,.
\end{equation}

The saddle point $N_{s+}$ resides precisely at the edge of the region where no finite action perturbative solutions exist. Since the integration contour for the background passes through this point, it makes sense to deform the contour ever so slightly away from the excluded half-line $N>N_{s+}, (N \in \mathbb{R}),$ and evaluate the action at
\begin{align}
N^2 = N_{s+}^2 + \delta^2\,,
\end{align}
where $\delta$ is a small complex number. Then the action can be evaluated by expanding in $\delta,$ \textit{e.g.}
\begin{align}
\left[ \frac{(1-\beta)\alpha}{(1-\alpha)\beta} \right]^{\mu/2}\bigg|_{\sqrt{N_{s+}^2+\delta^2}} & \rightarrow \, e^{-ik(\eta_0 + \eta_1)} \left( 1-\frac{i \delta k N_{s+}H^{3/2}}{(q_0 q_1)^{3/4}}\right)\,, 
\end{align}
and subsequently taking the limit $\delta \rightarrow 0$.
With this prescription, we evaluate the classical action $\bar{S}^{(2)}$ evaluated at both saddle points,
\begin{align}
\bar{S}^{(2)}|_{N_{s\pm}}&=\frac{k^2}{2H^2}\frac{num}{ 
e^{ik(\eta_0\pm\eta_1)}(i k \eta_0 -1)(i k \eta_1\mp 1)
-e^{-ik(\eta_0\pm \eta_1)}(ik\eta_0+1)(ik\eta_1 \pm 1)
}
\label{eq:actionPrecise}
\end{align}
with the numerator
\begin{align}
num&=-4 i k \phi_0 \phi_1\\
& \mp e^{i k(\eta_0 \pm \eta_1)} \left[ \frac{\phi_0^2}{\eta_0}(ik\eta_1\mp 1) + \frac{\phi_1^2}{\eta_1}(ik \eta_0 -1)\right]\\
& \mp e^{-ik(\eta_0 \pm \eta_1)} \left[ \frac{\phi_0^2}{\eta_0}(ik\eta_1 \pm 1) + \frac{\phi_1^2}{\eta_1}(i k \eta_0 +1)\right]\,.
\end{align} 
Note that equation \eqref{eq:actionPrecise} is a precise form of equation \eqref{actionzeroq0} evaluated in $N_{s\pm}$ when approached from above and below the branch-cut.

This finally enables us to take the limit of vanishing initial scale factor, used in selecting the Bunch-Davies vacuum. This is equivalent to the limit where $\eta_0$ approaches minus infinity since the scale factor and conformal time are related by $q=1/(H\eta)^2$. As is clear from the above expressions, the classical action does not converge in the limit of $\eta_0\to -\infty$ along the real axis. One can regularise the limit by adding a (vanishingly) small imaginary part to the initial scale factor. This however leads to two inequivalent results. In the limit $\eta_0\to - \infty (1-i \epsilon)$ for positive $\epsilon$ or equivalently $q_0 \rightarrow 0$ with $\text{Im } q_0 > 0$ -- normally considered for the Bunch-Davies calculation -- the classical action reduces to
\begin{align}
\bar{S}^{(2)}|_{N_{s\pm}}&\to 
\frac{k^2}{2H^2}
\frac{
\mp e^{-i k(\eta_0 \pm \eta_1)} \left[ \frac{\phi_0^2}{\eta_0}(ik\eta_1\pm 1) + \frac{\phi_1^2}{\eta_1}(ik \eta_0 +1)\right]
}
{
(-1)e^{-ik(\eta_0\pm\eta_1)}(i k \eta_0 + 1)(i k \eta_1\pm 1)
}\\
&=\frac{k^2\phi_1^2}{2H^2\eta_1 (\pm i k \eta_1 + 1)}\\
&=\frac{q_1k^2\phi_1^2}{2 (\pm i k - H\sqrt{q_1})}\\
& \approx -\frac{\sqrt{q_1}k^2}{2H}\phi_1^2 \mp i \frac{k^3}{2H^2} \phi_1^2 \,.
\end{align}
Since the semi-classical approximation amounts to exponentiating the classical action, \textit{i.e.} $e^{i \bar{S}^{(2)}/\hbar}$, we observe that the saddle point $N_{s-}$ corresponds to a Gaussian while $N_{s+}$ corresponds to a non-normalisable, inverse Gaussian, mode. Meanwhile, taking the limit with negative $\epsilon$ or equivalently with $\text{Im } q_0 < 0$ instead, we obtain the complex conjugate result. 
The saddle point $N_{s+}$ then corresponds to Gaussian and $N_{s-}$ to inverse Gaussian fluctuations. In either case we observe that the fluctuations are always stable at one saddle point, and unstable at the other.

\begin{figure}[h] 
\centering
\begin{minipage}{0.45\textwidth}
\includegraphics[width=\linewidth]{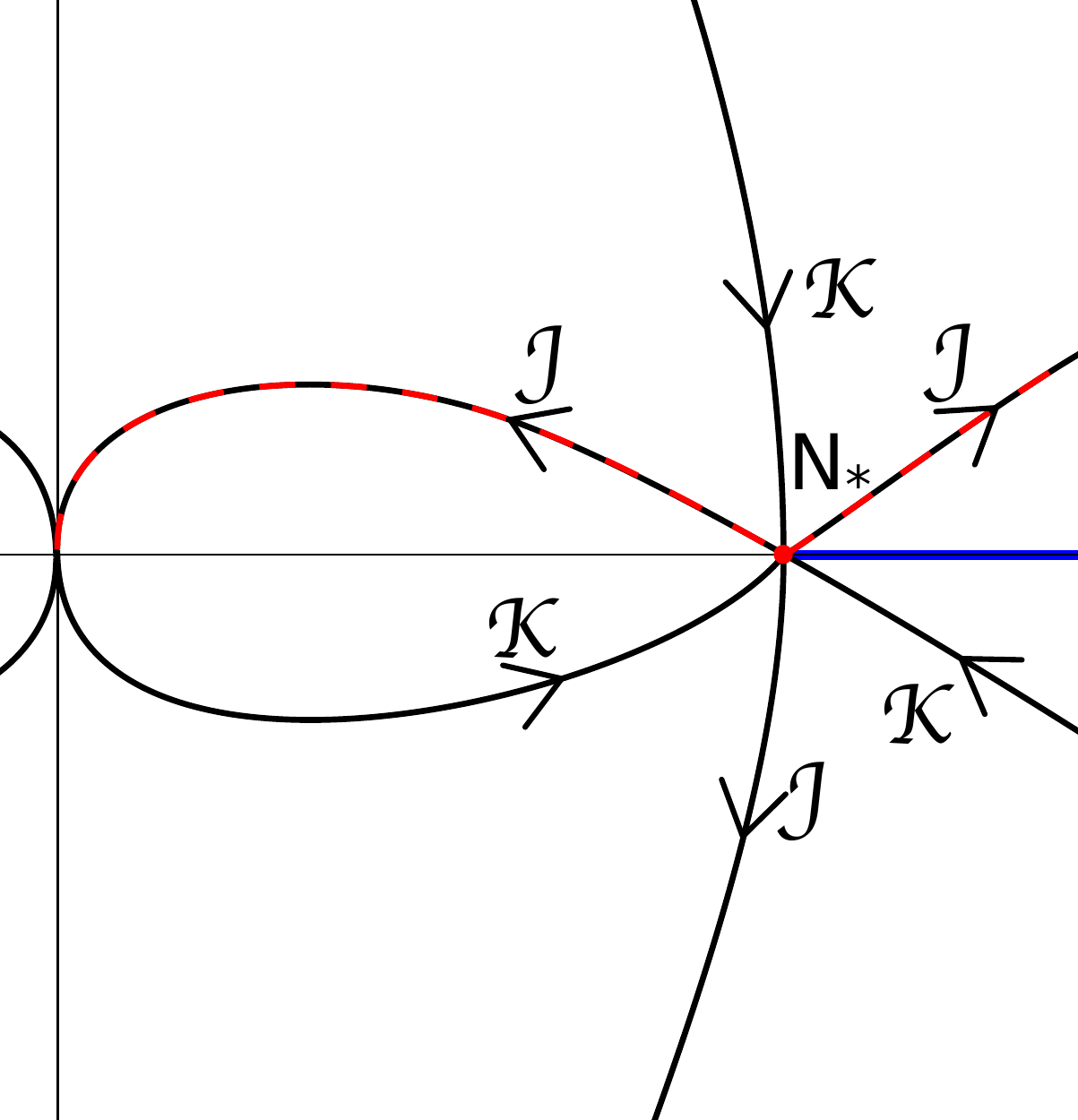}
\end{minipage}~
\begin{minipage}{0.45\textwidth}
\includegraphics[width=\linewidth]{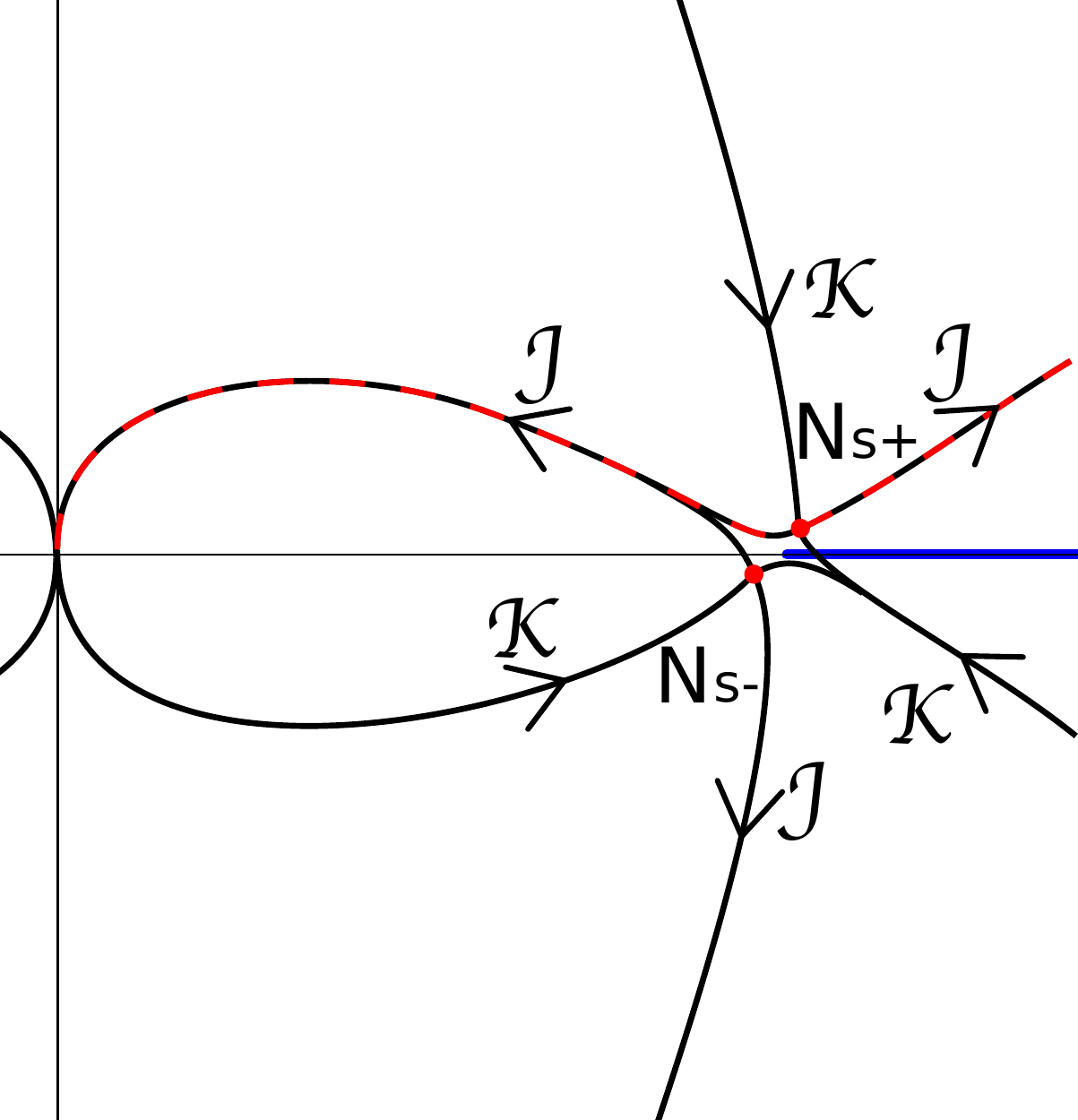}
\end{minipage}
\caption{Picard-Lefschetz theory for the lapse integral in the limit $q_0\to 0$. The dark lines show the lines of steepest ascent/descent indicated by $\mathcal{K}$ and $\mathcal{J}$. Using Picard-Lefschetz theory the real half-line $(0^+,\infty)$ is deformed to the red dashed curves. Left: the Lefschetz thimbles for degenerate saddle point $N_*$ when setting $q_0=0$. Right: the Lefschetz thimbles in the limit $q_0\to 0$ approached from the upper half plane $\text{Im } q_0 >0$. We observe that when approaching $q_0=0$ from above, the saddle point $N_{s+}$ is relevant while the saddle point $N_{s-}$ is irrelevant. When approaching $q_0$ from below the saddle points switch positions.}
\label{fig:Perturb}
\end{figure}

In order to determine the relevant saddle points in the saddle point approximation, we apply Picard-Lefschetz theory to the limits $q_0\to 0$ with $\text{Im } q_0 <0$ and $\text{Im } q_0 >0$ (see figure \ref{fig:Perturb}). We observe that the saddle points move in the complex plane. For the limit $\text{Im } q_0>0$ we observe that the lines of steepest ascent of the $N_{s+}$ saddle point intersect the positive half line while the lines of steepest ascent of $N_{s-}$ curve away from the real line. We thus conclude that only $N_{s+}$ is relevant in this limit. In the limit $\text{Im } q_0<0$ the saddle points switch making $N_{s-}$ relevant. We thus observe that the propagator always selects the \emph{unstable} saddle point that contributes to the Feynman propagator, while the stable saddle point is irrelevant to the propagator in the limit of zero initial scale factor. We conclude that the semi-classical description of an initial (quasi-) de Sitter phase does not lead to the Bunch-Davies vacuum often assumed in inflation when studied as a QFT on curved spacetime. Instead the Feynman propagator selects the unsuppressed (inverse Gaussian) fluctuations. Before discussing the implications of this result, it is useful to verify it by relating it to a similar calculation arising for the no-boundary proposal.

\section{Flat space limit of the no-boundary proposal} \label{sec:nb}

An alternative manner in which to study the early time limit of de Sitter space in the flat slicing is to set $q_0=0$ from the start and let the spatial curvature parameter $K$ be positive at first, and then approach zero. This effectively corresponds to the flat space limit of the no-boundary proposal, in which one sums over compact metrics between an initial point $q_0=0$ and a final 3-surface specified by $q_1$. The relevant calculation has been performed in the papers \cite{Feldbrugge:2017kzv,Feldbrugge:2017fcc,Feldbrugge:2017mbc} which we refer to for additional details. To include positive spatial curvature, we generalise the metric to read $\mathrm{d}s^2=-\frac{N^2}{q}\mathrm{d}t^2 + q \mathrm{d} \Omega^2,$ where $\mathrm{d}\Omega^2$ is the metric on a 3-sphere of curvature $6K.$ Taking the limit of zero spatial curvature can then be thought of as enlarging the sphere to reduce the local curvature, and in the infinite size limit where flat space is reached we assume that the volume is suitably regularised (\textit{e.g.} by imagining that the flat 3-space has the topology of a torus). Alternatively, one may take the limit where at fixed $K$ the final scale factor $q_1$ tends to infinity, as in this limit the spatial curvature also becomes insignificant. For the purposes of illustration, we will take the point of view that $K$ is reduced with $q_1$ held fixed.

\begin{figure}
\begin{minipage}{150pt}
		\includegraphics[width=0.9\linewidth]{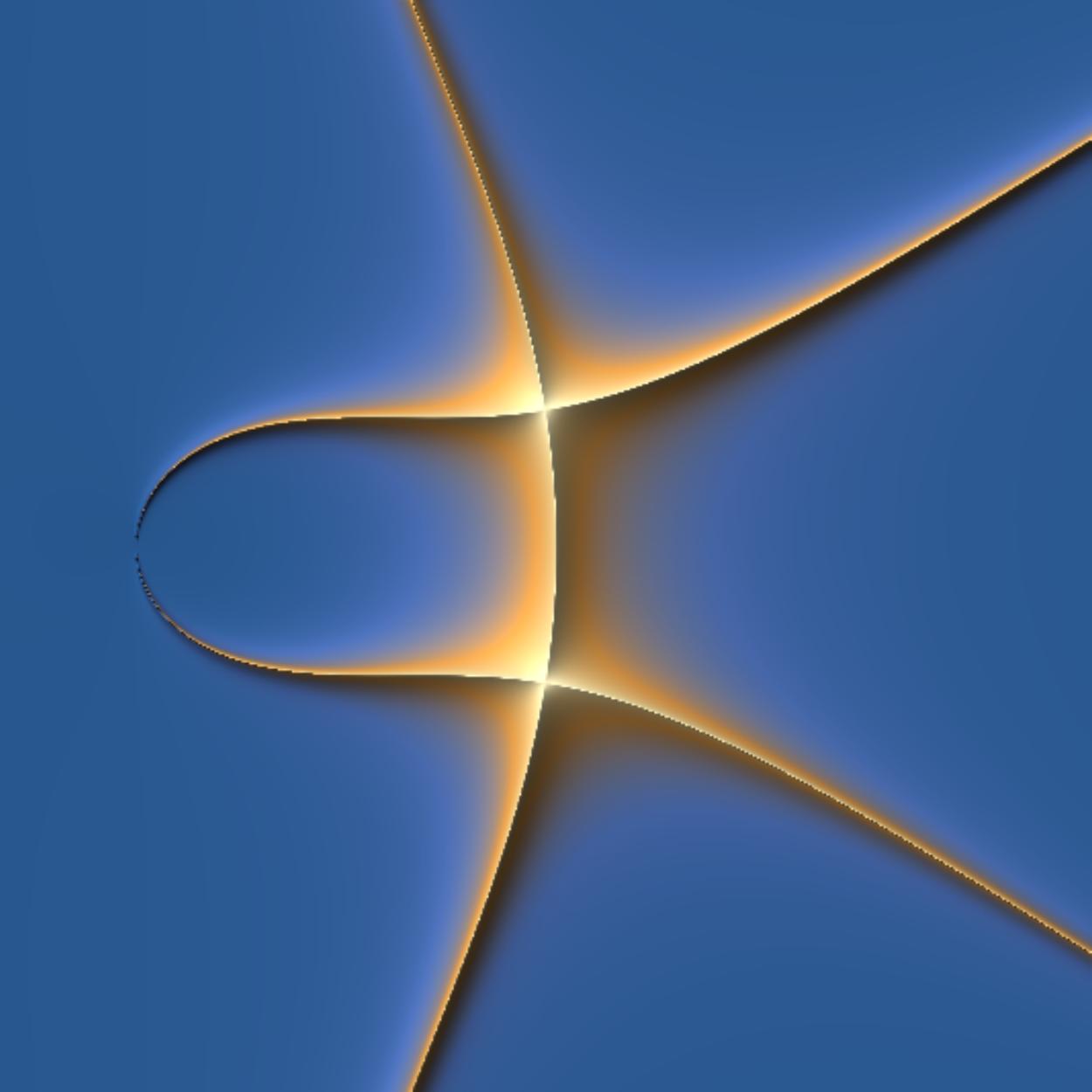}
	\end{minipage}%
	\begin{minipage}{150pt}
		\includegraphics[width=0.9\linewidth]{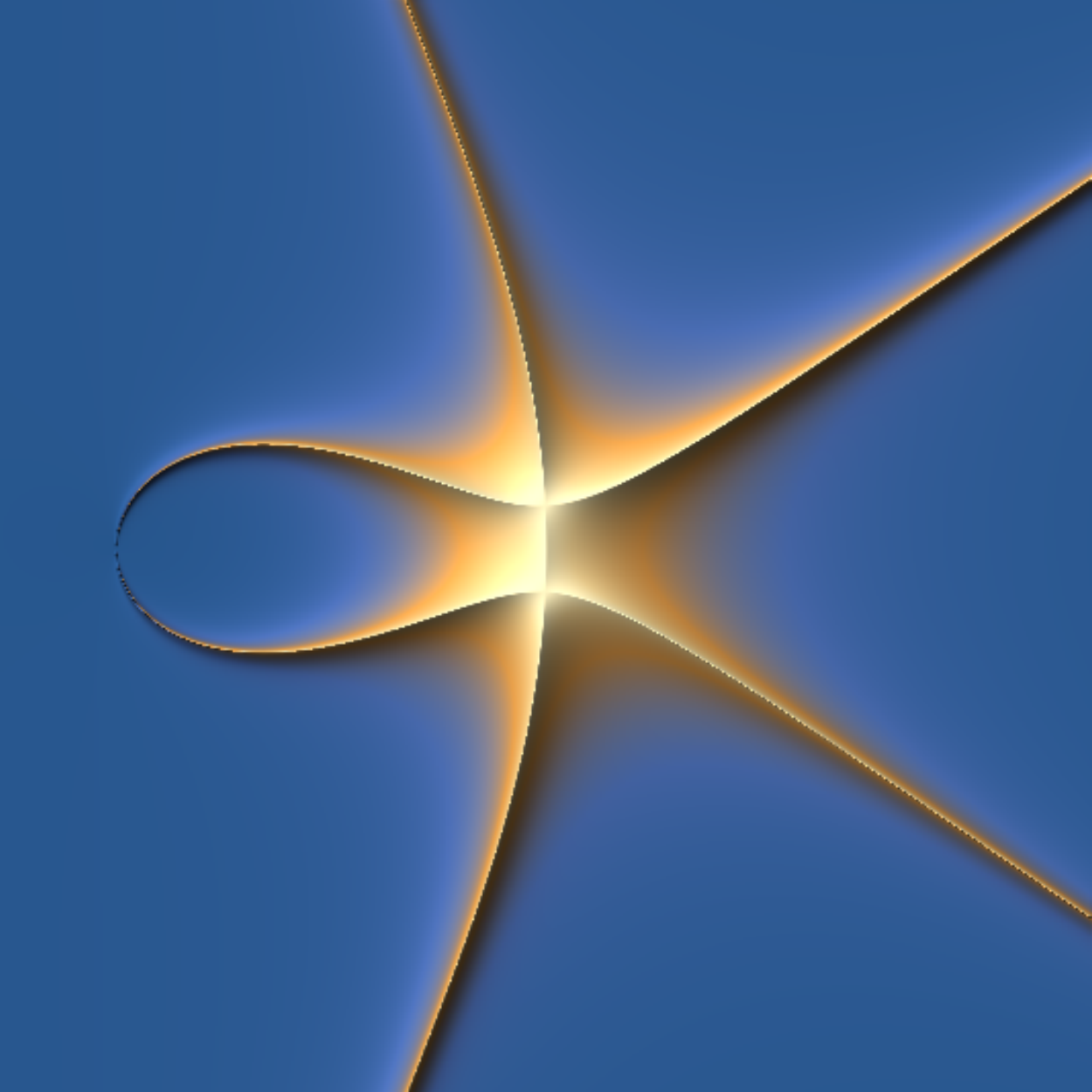}
	\end{minipage}%
	\begin{minipage}{150pt}
		\includegraphics[width=0.9\linewidth]{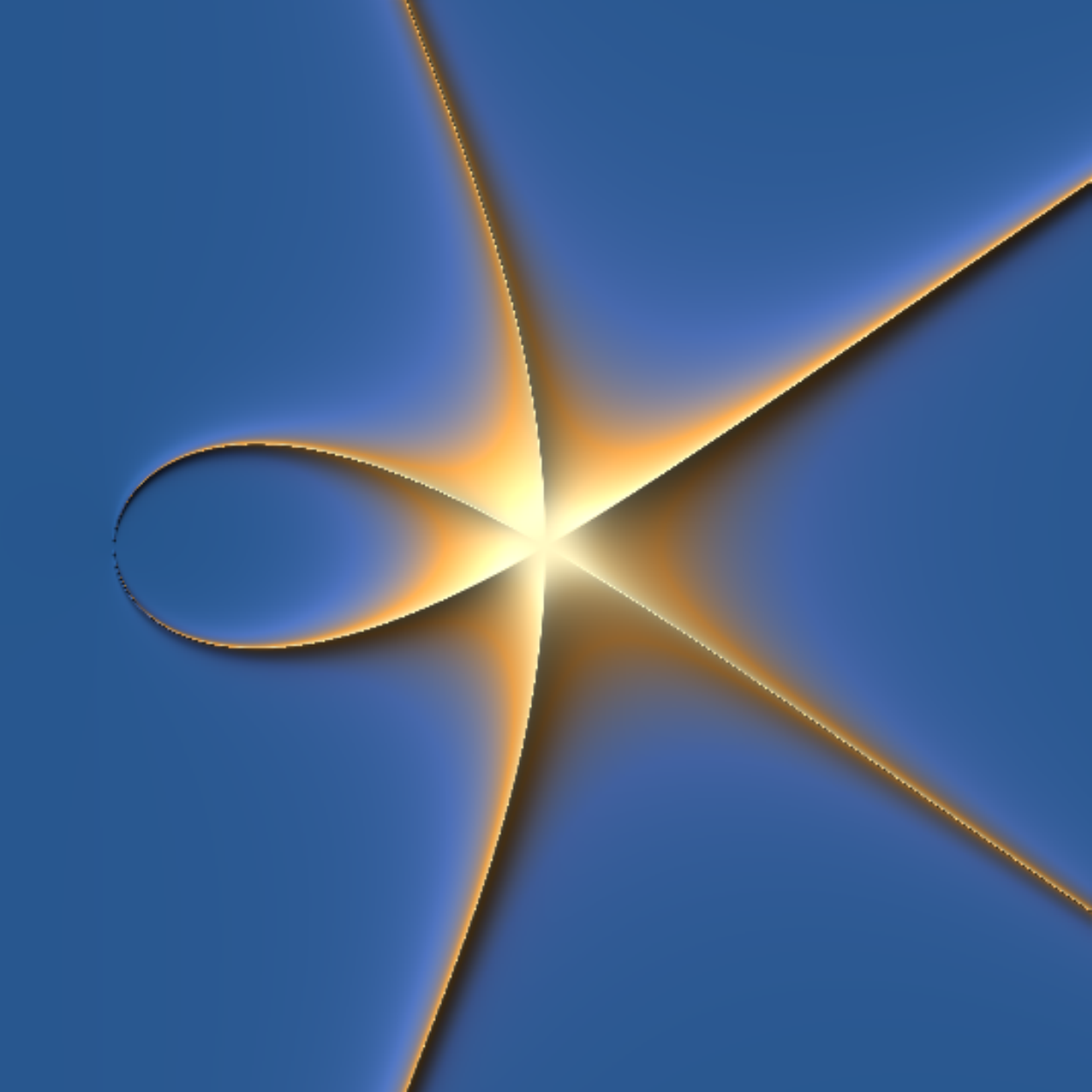}
	\end{minipage}%
	\caption{Picard Lefschetz theory in $N$ with $q_0=0,$ in the limit $K \to 0,$ with $\Lambda=3$ and $q_1=10$. The curves are the lines of steepest ascent and descent emanating from the saddle point at the crossings of the lines. The origin the point where the two lines meet at the left. Note that the figure is symmetric around the real line. {\it Left panel:} Spatial curvature $K=1$: this is the standard no-boundary graph. {\it Middle panel:} Spatial curvature $K=1/10$.  {\it Right panel:} Spatial curvature $K=0$.}
	\protect
	\label{fig:thimble}
\end{figure}

In the presence of spatial curvature, the saddle points are qualitatively different from the flat case in that they reside at complex values of the lapse function,
\begin{align}
N_s &= \pm \frac{1}{H^2} \left[ (H^2 q_1 - K)^{1/2} \pm i K^{1/2}\right]\,.
\end{align}
As the curvature $K$ is reduced, the saddle points approach the real $N$ line, and in the limit of zero curvature they reach it at $\pm N_\star = \pm \frac{\sqrt{q_1}}{H}.$ Note that $N_\star$ also happens to be the saddle point of the perturbative action \eqref{actionzeroq0}. Fig. \ref{fig:thimble} illustrates the positions of the saddle points, and the corresponding lines of steepest ascent and descent of the Morse function $h=\text{Re}(iS)$.  In the limit of $K=0$ a degenerate saddle point of order 2 is obtained.

The form of the action \eqref{Nintegral} makes it clear that the integrand $e^{iS}$ is convergent near the origin ($N=0$) and at large $N$ in the regions just above the real $N$ line. Thus the integration contour of steepest descent from $N_\star$ resides entirely in the upper half plane\footnote{The steepest descent line emanating from $N_\star$ and approaching $-i\infty$ is traversed in both directions, so that its contribution to the integral cancels out.}. Then in the Lorentzian integral we are forced to approach $N_\star$ from above, where in the large $k$ limit the fluctuation action \eqref{actionzeroq0} approaches
\begin{equation}
\bar{S}^{(2)}(N_\star^+)  \rightarrow - \frac{i}{2} k q_1 \phi_1^2\,,
\end{equation}
implying a weighting $e^{+k q_1 \phi_1^2/(2\hbar)},$ where $N_\star^+$ denotes $N_\star$ approached from above. Note that the action at $N_\star^+$ is independent of $\Lambda,$ and scales as $k \phi_1^2$ rather than $k^3 \phi_1^2/\Lambda$. The large $k$ limit corresponds to modes with short wavelengths, which are not yet frozen.  Another interesting limit is that of large final scale factor $\sqrt{q_1}$, where
\begin{equation}
\bar{S}^{(2)}(N_\star^+)  \rightarrow  - \frac{\sqrt{q_1}}{2H} k^2 \phi_1^2 -  \frac{i}{2H^2} k^3 \phi_1^2\,. \label{Sabovecut}
\end{equation}
The frozen modes acquire a scale-invariant \emph{inverse} Gaussian distribution, with weighting $e^{+k^3 \phi_1^2/(2H^2 \hbar)}$. This confirms the conclusion that a phase of de Sitter (or quasi-de Sitter) expansion considered all the way into our past does not lead to the Bunch-Davies vacuum, and hence cannot by itself explain the origin of structure in our universe.

\section{Inflatable initial conditions} \label{InitialConditions}

The discussion above indicates that, in a consistent semiclassical treatment of both the background and the perturbations, sending the initial size of the universe to zero in a well-defined way results in an unacceptable, unbounded probability distribution for the perturbations.  One may, however, wonder whether there might be some other way to specify the initial conditions for inflation in a natural manner, which would avoid the interference of backgrounds described above and would be able to recover the limit of QFT in curved spacetime. 

In Ref.~\cite{Feldbrugge:2019}, we discuss the use of localized initial and final quantum states in relativistic, diffeomorphism invariant (and hence constrained) quantum mechanics. We show there that, in order to prescribe states which are localized, it is necessary to employ ``off-shell" wavefunctions, namely, wavefunctions which are {\it not} annihilated by the Hamiltonian constraint. These states represent the quantum amplitude resulting from a preparation process and {\it not} the dynamics of the system itself, hence the system Hamiltonian alone does not annihilate the state. Once such states are evolved with the Feynman propagator, outside of the preparation region they become a superpositon of ``on-shell'' physical states of the system, independent of the preparation device. In the present section we shall make use of this formalism to define an inflating initial state which avoids the problem of multiple contributing backgrounds. The initial state will be localized in superspace at small scale factor, and endowed with a positive expansion velocity. In the localized region of phase space in which it is initially prepared, the wavefunction does not satisfy the Hamiltonian constraint (or Wheeler-DeWitt equation). However, once this initial wavefunction is propagated to large scale factor, it becomes that for a large, expanding universe and satisfies the Wheeler-DeWitt equation.  

In section \ref{sec:background}, we obtained the Feynman propagator for the background as an ordinary integral over $N$, \textit{i.e.} 
\begin{align}
G[q_1;q_0] = \sqrt{\frac{3V_3 i}{4\pi \hbar}} \int_{0^+}^\infty \frac{\mathrm{d}N}{\sqrt{N}}  e^{\frac{i}{\hbar}\bar{S}^{(0)}[q_1;q_0;N]}\,, \label{Nintegral2}
\end{align}
in terms of the classical action
\begin{align}
\bar{S}^{(0)}[q_1;q_0;N] = V_3 \left[ \frac{\Lambda^2}{36}N^3 -\frac{\Lambda}{2}(q_0+q_1)N -\frac{3(q_1-q_0)^2}{4N}\right]\,. \label{classicalaction}
\end{align}
Implicitly, this calculation assumed that the universe was initially in a position eigenstate, with the scale factor squared $q$ being precisely equal to $q_0.$ As a direct consequence of the uncertainty principle, the initial momentum of the universe was thus maximally unknown -- one may interpret the resulting interference of an expanding and a bouncing solution as a reflection of this fact. We can now consider a more general initial state $\psi_0,$ which is evolved  by convolution with the propagator,
\begin{align}
G[q_1;\psi_0] = \int G[q_1;q_0] \psi_0(q_0) \mathrm{d}q_0 = \int_{0^+}^\infty \int G[q_1;q_0;N] \psi_0(q_0) \mathrm{d}q_0 \mathrm{d}N\,.
\end{align}
Here we will consider the choice 
\begin{align} \label{Initialstate}
\psi_0(q_0) = \frac{1}{\sqrt[4]{2\pi \sigma^2}}e^{\frac{i}{\hbar} p(q_i) q_0 - \frac{(q_0-q_i)^2}{4 \sigma^2}}\,,
\end{align}
which expresses the idea that the universe starts with a certain size $q_i$ (with uncertainty $\sigma$), and that we choose the momentum $p(q_i)= \pm V_3 \sqrt{3\Lambda q_i}$ so that the universe is initially contracting (positive sign in $p(q_i)$) or expanding (negative sign in $p(q_i)$). Below we will specialize to the expanding case. Note that this state simply represents a {\it choice}, not explained by inflation. Our strategy here is rather to imagine the physical situation in which we know with some confidence that the universe is expanding, and is likely to be of a certain size already, and then explore the consequences of this assumed initial state. The mathematical form of the initial state is that of a generalized coherent state, which distributes the uncertainty between ``position'' $q$ and momentum $p$ depending on the value of the spread $\sigma.$ (We have performed analogous calculations with different phase correlations in $\psi_0(q_0),$ in particular with the momentum factor $e^{ip(q_0)q_0/\hbar}$ instead of $e^{ip(q_i)q_0/\hbar},$ and have checked that qualitatively similar results are obtained. Similarly, one may consider states which have support only at positive $q_0$, again with analogous results. However, the present choice is technically the simplest.)

The convolution of the propagator with the state \eqref{Initialstate} is Gaussian and can thus be evaluated exactly, yielding
\begin{align}
\int G[q_1;q_0;N] \psi_0(q_0) \mathrm{d}q_0 = 
 \frac{1}{\sqrt[4]{2\pi}}
  \sqrt{\frac{i 3 V_3 \sigma N}{\hbar N + 3 i V_3 \sigma^2}}
e^{\frac{i}{\hbar} \bar{S}^{(0)}[q_1;\bar{q}_0;N]+ \frac{i}{\hbar} p(q_i) \bar{q}_0 - \frac{(\bar{q}_0-q_i)^2}{4 \sigma^2}}
\end{align}
with
\begin{align}
\bar{q}_0=
\frac{N q_i - i (N^2 \Lambda V_3 -2 N p(q_i) - 3 q_1 V_3) \sigma^2/\hbar}{N + 3 i \sigma^2 V_3 /\hbar}\,. \label{barq0}
\end{align}
Note that in the limit $\sigma \to 0$, we localize $\bar{q}_0 \to q_i$.

\begin{figure}[h] 
\centering
\begin{minipage}{0.49\textwidth}
\includegraphics[width=\linewidth]{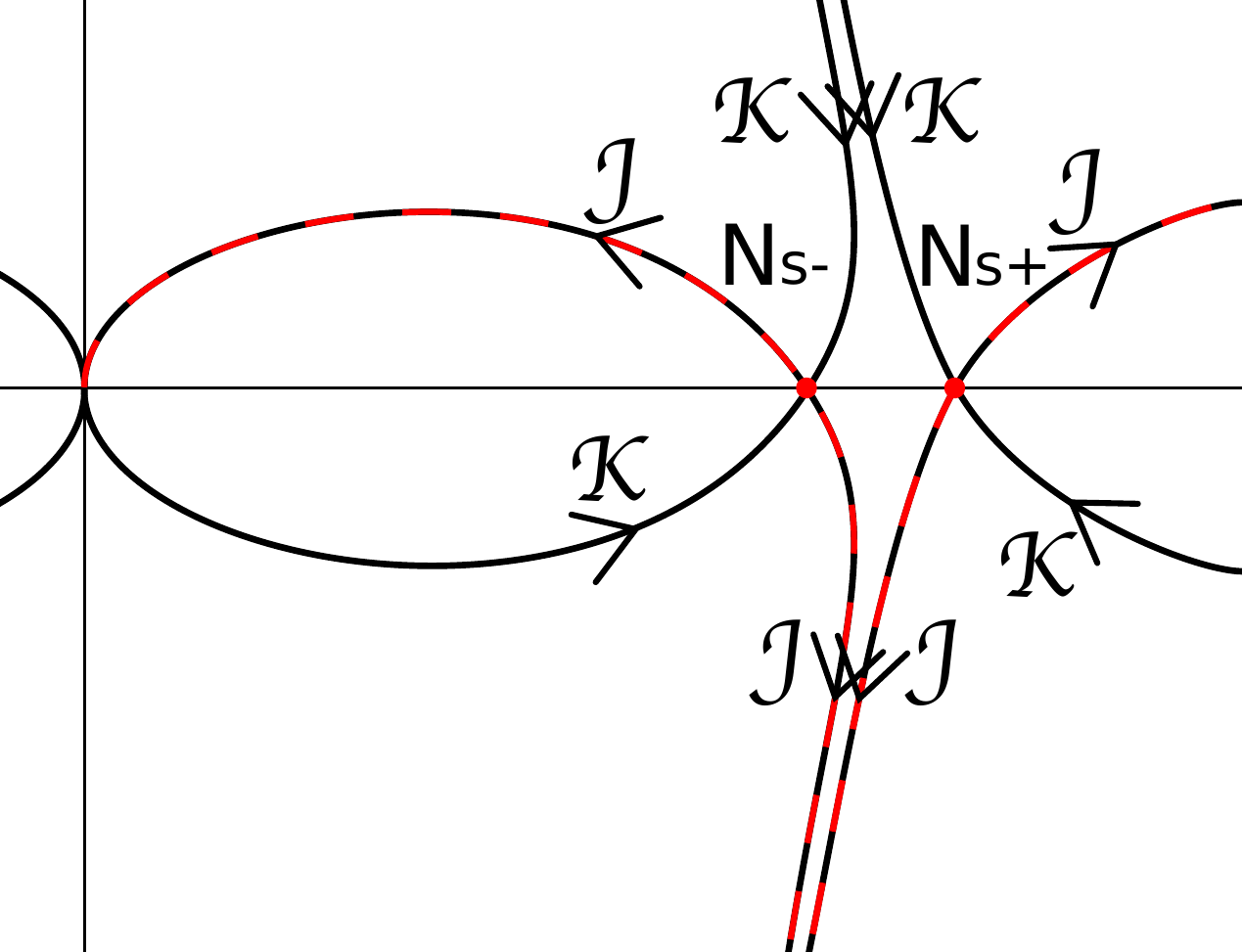}
\end{minipage}~
\begin{minipage}{0.49\textwidth}
\includegraphics[width=\linewidth]{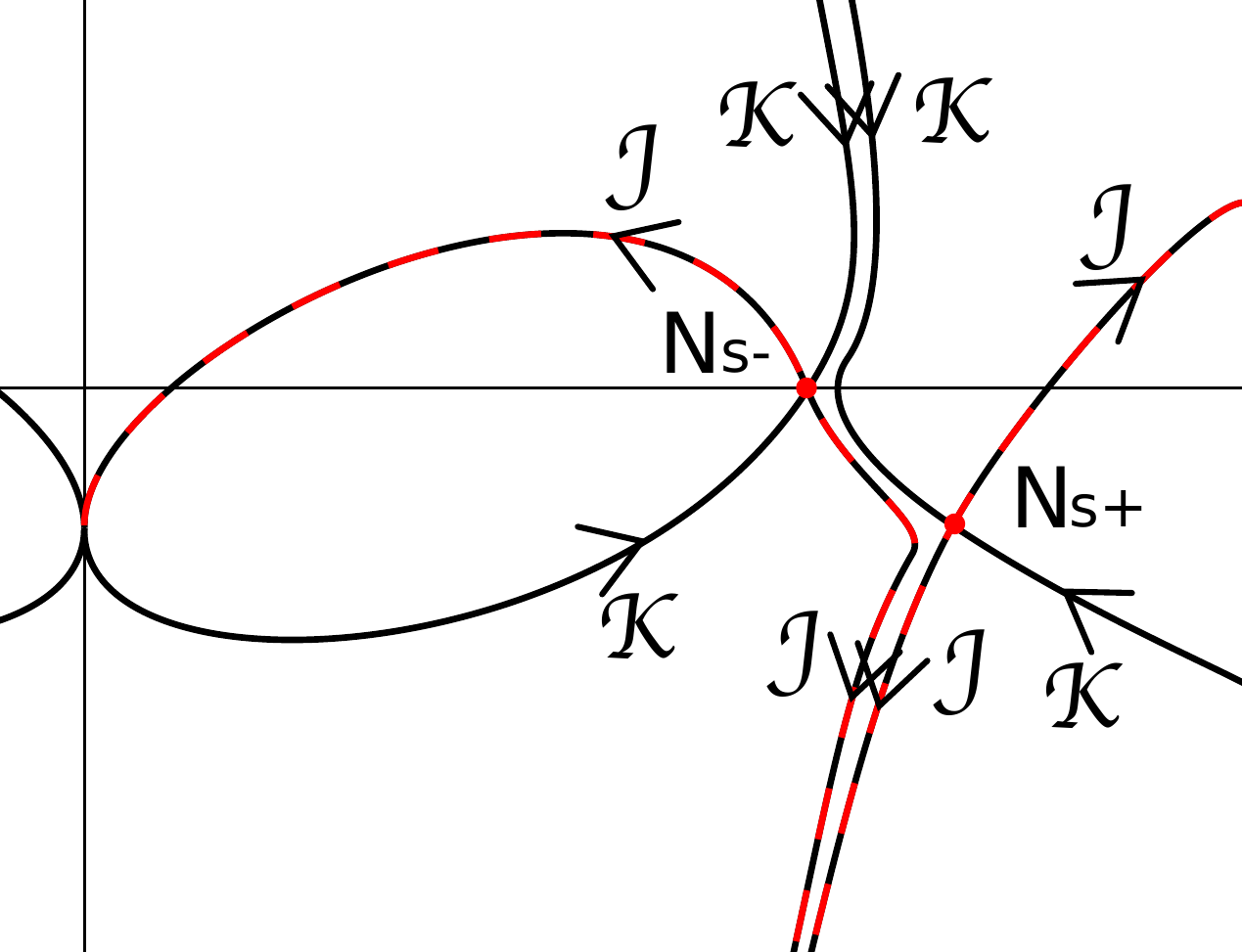}
\end{minipage}\\
\begin{minipage}{0.49\textwidth}
\includegraphics[width=\linewidth]{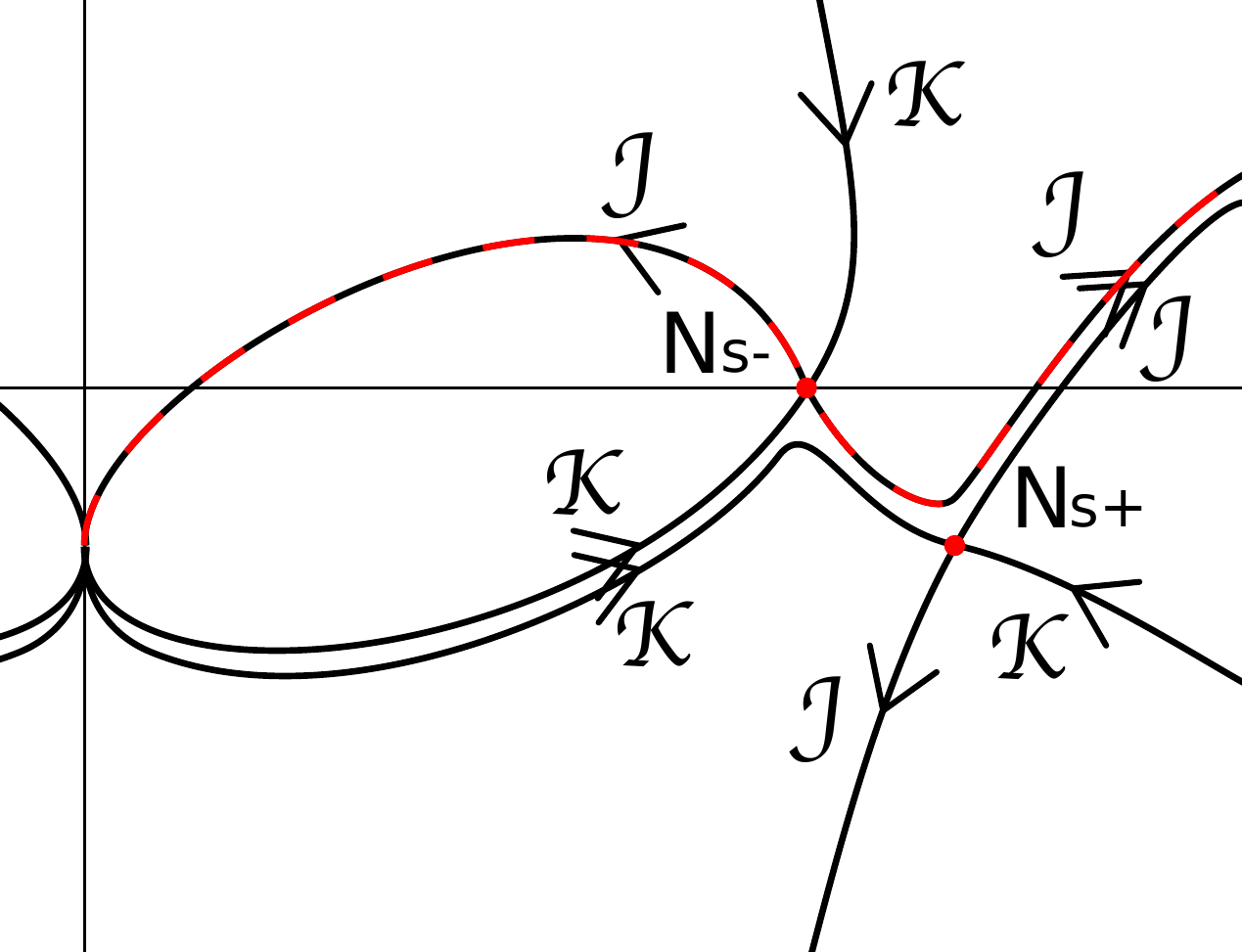}
\end{minipage}~
\begin{minipage}{0.49\textwidth}
\includegraphics[width=\linewidth]{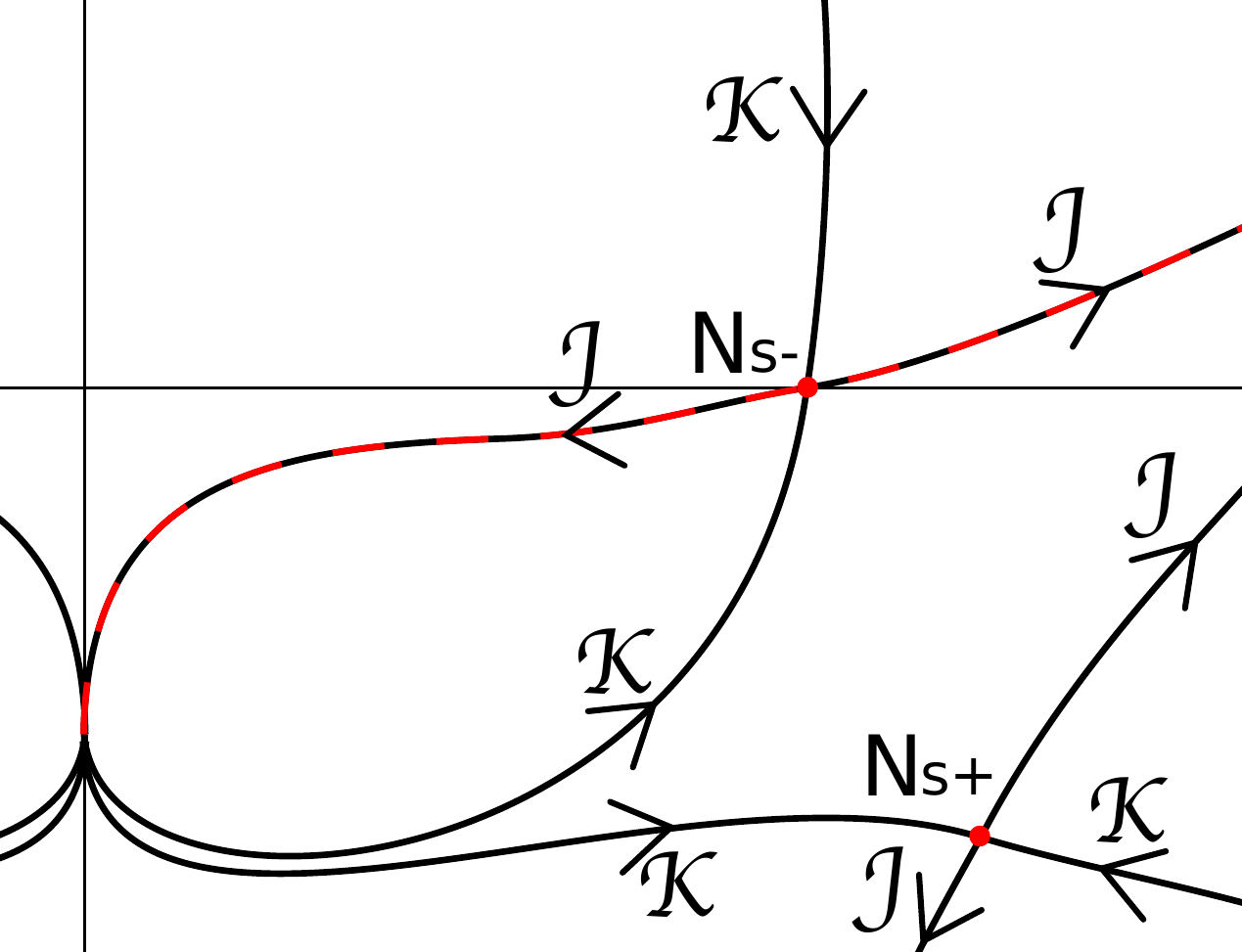}
\end{minipage}
\caption{The lines of steepest ascent and descent for the integral over $N$ as a function of an increasing initial localization $\sigma$ (the localization $\sigma$ increases from top left to top right, then lower left and finally lower right). When $\sigma$ becomes larger than the critical value $\sigma_c$ (which occurs in the transition from the top right to the lower left graph), a Stokes phenomenon happens: the line of steepest ascent $\mathcal{K}$ from the saddle point $N_{s+}=N^\sigma_{1,1}$ corresponding to the bouncing solution no longer intersects the original integration contour on the real $N$ line. Beyond this critical value, only one saddle point (namely the purely expanding saddle point $N_{s-}=N^\sigma_{1,-1}$) remains relevant to the semi-classical path integral and moreover this saddle point resides on the real $N$ line, so that it describes classical evolution. At this point standard quantum field theory in curved spacetime is recovered.}
\label{fig:PLsigma}
\end{figure}

The subsequent integral over $N$ can be evaluated using a saddle point approximation, with the help of Picard-Lefschetz theory. There are four saddle points of $\bar{S}[q_1;\bar{q}_0;N]$ in $N$ located at
\begin{align}
N^\sigma_{c_1,c_2}  &=  -\frac{3 i \sigma^2 V_3}{\hbar} +c_1 \sqrt{ \frac{3}{\Lambda}}\left(\sqrt{q_1} + c_2 \sqrt{q_i + \frac{2 i p(q_i) \sigma^2}{\hbar} - \frac{3 \Lambda \sigma^4 V_3^2}{\hbar^2}} \right) \\ &= \sqrt{ \frac{3}{\Lambda}} \left( c_1 \sqrt{q_1}+ c_2\sqrt{q_i}\right)  +(\mp c_2-1) \frac{3 i \sigma^2 V_3}{\hbar} \,, \label{Nssigma}
\end{align}
with $c_1,c_2=\pm 1$ and where we have used $p(q_i) = \mp V_3 \sqrt{3\Lambda q_i}$ for the initially expanding/contracting cases respectively. The corresponding Lefschetz thimbles for the boundary conditions $0<q_i \ll q_1$ are given in Fig. \ref{fig:PLsigma}, for the case of an expanding initial state. 

In the limit of zero spread $\sigma =0$ (the upper left panel of Fig. \ref{fig:PLsigma}), we saw that $\bar{q}_0 \to q_i$ and thus we recover our earlier results, namely that the propagator is dominated by both an expanding and a bouncing solution. The expanding solution is given by $(c_1,c_2)=(+1,-1)$, while the bouncing solution is given by $(c_1,c_2)=(+1,+1),$ and we will keep referring to these saddle points as $N_{s-}=N^\sigma_{1,-1}$ and $N_{s+}=N^\sigma_{1,1}$ respectively, by analogy with our discussion in earlier sections. As $\sigma$ is increased, the saddle point $N_{s-}$ corresponding to expansion stays put, and in fact for this saddle point $\bar{q}_0=q_i,$ that is to say the expanding solution starts from the central value of $q.$ Meanwhile, the saddle point $N_{s+}$ corresponding to a bounce starts to move off the real axis. For sufficiently small $\sigma$, both saddle points are relevant (top right panel). However, as $\sigma$ reaches a critical value $\sigma_c$ we observe a Stokes phenomenon after which solely the expanding solution is relevant for the Feynman propagator (see the lower left panel of Fig. \ref{fig:PLsigma} for an illustration). Thus, for a localization with $\sigma > \sigma_c,$ the propagator is dominated by a single saddle point, which moreover resides on the real $N$-axis, implying that the treatment of both the background and the fluctuations will be well approximated by quantum field theory on curved space-time \footnote{A similar construction, formulated in terms of Robin boundary conditions and using some of the insights presented in this paper, was recently implemented for the no-boundary proposal in \cite{DiTucci:2019dji}.}. If we \emph{assume} the appropriate Bunch-Davies state for the perturbations of the (expanding) initial state, we will recover the predictions of inflation, as long as the perturbative action remains stable and well-defined along the entirety of the thimble. We will explore this issue in the next section. Note however that in this framework inflation does not explain the Bunch-Davies state by itself -- rather it must be put in by hand from the outset and eventually an additional theory will be required to explain it. 

The critical localization $\sigma_c$ can be found by solving for the condition that the two saddle points are linked by a line of steepest ascent/descent,
\begin{align}
\text{Im}\left[i\bar{S}^{(0)}+ip(q_i)\bar{q}_0-\frac{\hbar(\bar{q}_0-q_i)^2}{4\sigma^2}\right]_{N_{s-}}=\text{Im}\left[i\bar{S}^{(0)}+ip(q_i)\bar{q}_0-\frac{\hbar(\bar{q}_0-q_i)^2}{4\sigma^2}\right]_{N_{s+}}\,. \label{Stokes}
\end{align}
The action at ${N}_{s-}$ is very simple (and real), and is given by
\begin{equation}
\begin{split}
\left[\bar{S}^{(0)}+p(q_i)\bar{q}_0+i\frac{\hbar(\bar{q}_0-q_i)^2}{4\sigma^2}\right]_{N_{s-}}= - V_3 \sqrt{\frac{\Lambda}{3}} \left( 2 q_1^{3/2} + q_i^{3/2}\right)\,.
\end{split}
\label{actionnob}
\end{equation}
For the bouncing saddle point the action reads
\begin{equation}
\begin{split}
\left[\bar{S}^{(0)}+p(q_i)\bar{q}_0+i\frac{\hbar(\bar{q}_0-q_i)^2}{4\sigma^2}\right]_{N_{s+}} =  &- V_3 \sqrt{\frac{\Lambda}{3}} \left[2 q_1^{3/2} +  \sqrt{q_i} \Bigl(\frac{ 5 q_i \hbar^2 - 36 \Lambda \sigma^4 V_3^2 }{\hbar^2}  \Bigr) \right] + \\
& - i \left(\frac{ \Lambda \sigma^4  V_3^2}{\hbar^2} - q_i\right)12 \Lambda \sigma^2 \frac{V_3}{\hbar}
\end{split}
\end{equation}
Therefore, we find that the two saddles are linked by a Stokes line if
\begin{equation}
4 \frac{V_3}{\hbar^2} \sqrt{\frac{q_i \Lambda}{3}} (\hbar^2 q_i - 9 V_3^2 \Lambda \sigma^4) = 0\,,
\end{equation}
implying that the critical localization is given by 
\begin{align}
\sigma_c = \left( \frac{q_i \hbar^2}{9 \Lambda V_3^2} \right)^{1/4}\,. \label{sigapprox}
\end{align}
Note that the critical localization $\sigma_c$ is independent of the final state $q_1,$ but that it depends on the inflationary vacuum energy $\Lambda$.

\begin{figure}
\centering
\includegraphics[width=0.5\textwidth]{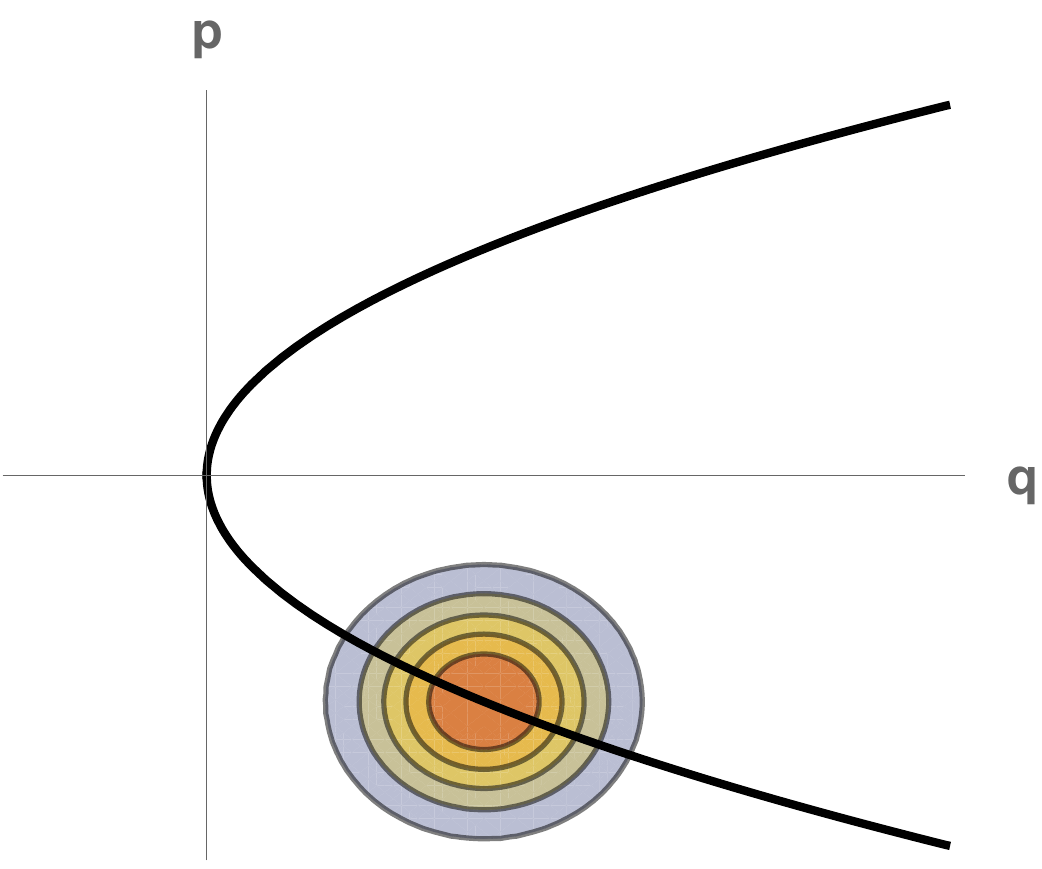}
\caption{The probability distribution of the initial wave function in phase-space $(q_i,p_i)$. The black line is the Hamiltonian constraint. The horizontal axis is $q_i$ and the vertical axis  $p_i$. In order to have only an expanding universe, the wave function needs to have support only in the lower right quadrant.}
\label{fig:wavefunction}
\end{figure}

We can set up initial conditions for an expanding universe whenever 
\begin{align}
q_i >  \sigma_c\,,
\end{align}
since this will allow us to specify the momentum $p(q_i)$ with sufficient accuracy, \textit{i.e.}
\begin{align}
|p(q_i)| > \frac{\hbar}{2 \sigma_c}\,.
\end{align}
This is confirmed by evaluating the Wigner function of the initial state 
\begin{align}
P(q,p) = \frac{1}{\pi \hbar} \int \psi_0^*(q+y)\psi_0(q-y) e^{2i p y/\hbar} \mathrm{d}y\,,
\end{align}
which is plotted in Fig. \ref{fig:wavefunction}. The size of the initial spatial slice $a_i$ depends on the value of the vacuum energy, as the condition $q_i > \sigma_c$ translates into
\begin{align}
a_i = \sqrt{q_i} > \frac{\hbar^{1/3}}{(9\Lambda)^{1/6}V_3^{1/3}} \qquad \quad \text{or} \qquad \quad a_i^3 V_3 > \frac{\hbar}{3 M_{Pl} \sqrt{\Lambda}},
\label{stokesbound}
\end{align}
where in the last expression we have restored $M_{Pl}$ by dimensions. This agrees with the bound we derived through a heuristic argument in Eq.~(\ref{estbd}) above.  

The implied bound means, for example, that for inflation at the grand unified scale $\Lambda \sim \left( 10^{-3}\right)^4 M_{Pl}^4,$ one can only describe the beginning of inflation in the context of QFT in curved spacetime when the initial size is larger than about a million Planck volumes, or about a hundred Planck lengths in linear size. Note that the current analysis is conservative, in that we have neglected any constraints that might arise when adding perturbations -- as we will see, the inclusion of perturbations leads to a small strengthening of the bound.

\subsection{Stable Perturbations} \label{sec:stableperts}

Having established that with a suitable initial state the background evolution can be consistently reduced to a configuration describing an expanding universe only, we would like to see if the perturbations are also well-behaved in this background. The effect of the convolution with the initial state is to substitute the initial scale factor $q_0$ with an ``effective'' initial size $\overline{q}_0$ defined as 
\begin{equation}
\overline{q}_0= \frac{N q_i - i ( 3 N^2 H^2 V_3 - 2 N p(q_i) - 3 q_1 V_3) \sigma^2/\hbar}{N + i 3 \sigma^2 V_3/\hbar}\,,
\end{equation}
with $p(q_i)= -  V_3 \sqrt{3 \Lambda q_i}$ and $H =\sqrt{ \frac{\Lambda}{3}}$. With this initial condition, the background solution becomes
\begin{equation}
\begin{split}
\bar{q}(t) = & H^2 N^2 t^2 + \frac{q_1 - \sqrt{q_i}(\sqrt{q_i} - i 6 H \sigma^2  V_3/\hbar) - H^2 N^2}{N + 3 i \sigma^2 V_3/\hbar} N t \\
& + \frac{N q_i + 3 i (q_1 - 2 H N \sqrt{q_i} - H^2 N^2)\sigma^2  V_3/\hbar}{N + 3 i \sigma^2 V_3/\hbar}\,.
\end{split}
\end{equation}
In analogy with the background solution that we obtained in the absence of an initial state, we may again write this as
\begin{equation}
\bar{q}(t) = H^2 N^2 (t - \alpha)(t - \beta)\,,
\label{qsol2}
\end{equation}
but this time with the $\sigma$-dependent coefficients
\begin{align}
\alpha &= \frac{ \left[ N^2 - (N^\sigma_{1,-1}N^\sigma_{1,1} + N^\sigma_{-1,1} N^\sigma_{-1,-1}) /2 + \sqrt{(N - N^\sigma_{1,-1})(N - N^\sigma_{1,1})(N - N^\sigma_{-1,1})(N - N^\sigma_{-1,-1})} \right] }{2 N (N + 3 i \sigma^2 V_3/\hbar)}\,, 
\label{eq:alphaGen}\\
\beta &= \frac{ \left[ N^2 - (N^\sigma_{1,-1}N^\sigma_{1,1} + N^\sigma_{-1,1} N^\sigma_{-1,-1}) /2- \sqrt{(N - N^\sigma_{1,-1})(N - N^\sigma_{1,1})(N - N^\sigma_{-1,1})(N - N^\sigma_{-1,-1})} \right] }{2 N (N + 3 i \sigma^2 V_3/\hbar)}\,.
\label{eq:betaGen}
\end{align}
Here the lapse values $N^\sigma_{c_1,c_2}$ with $c_1,c_2 = \pm 1$ correspond to the saddle points of the background action, determined earlier in Eq. \eqref{Nssigma}. Note that the definitions of $\alpha$ and $\beta$ are a simple generalization of the definitions in Eq. \eqref{ab}, the latter being recovered in the limit of $\sigma \to 0.$ It also remains the case that $\alpha = \beta$ at the saddle points.  

The equation of motion for the perturbations also remains identical in form, cf. Eq. \eqref{212}, but now with the $\sigma-$dependent coefficients $\alpha$ and $\beta$. The two linearly independent solutions are then once again $f(t)/\sqrt{q(t)}$ and $g(t)/\sqrt{q(t)}$ with
\begin{align}
f(t) &= \left[ \frac{t - \beta}{t  - \alpha } \right]^{\mu/2} \left[ (1 - \mu)(\alpha - \beta) + 2 (t - \alpha)\right]\,, \label{pertmode1}\\
g(t) &= \left[ \frac{t - \alpha}{t  - \beta } \right]^{\mu/2} \left[ (1 + \mu)(\alpha - \beta) + 2 (t - \alpha)\right]\,, \label{pertmode2}
\end{align}
where
\begin{equation}
\mu^2 = 1 - \left( \frac{2 k}{(\alpha - \beta) H^2 N} \right)^2\,.
\end{equation}
The two solutions correspond to the two square roots $\pm \mu$. At the (expanding) saddle point $N_{s-}=N^\sigma_{1,-1}$, which stays put on the real $N$ line as the localization $\sigma$ is varied, the solution given by $g(t)$ retains the form appropriate to the stable Bunch-Davies state. This is the solution that any initial state must single out in order to recover the standard description of inflationary fluctuations. 

\begin{figure}
\includegraphics[width=0.45\textwidth]{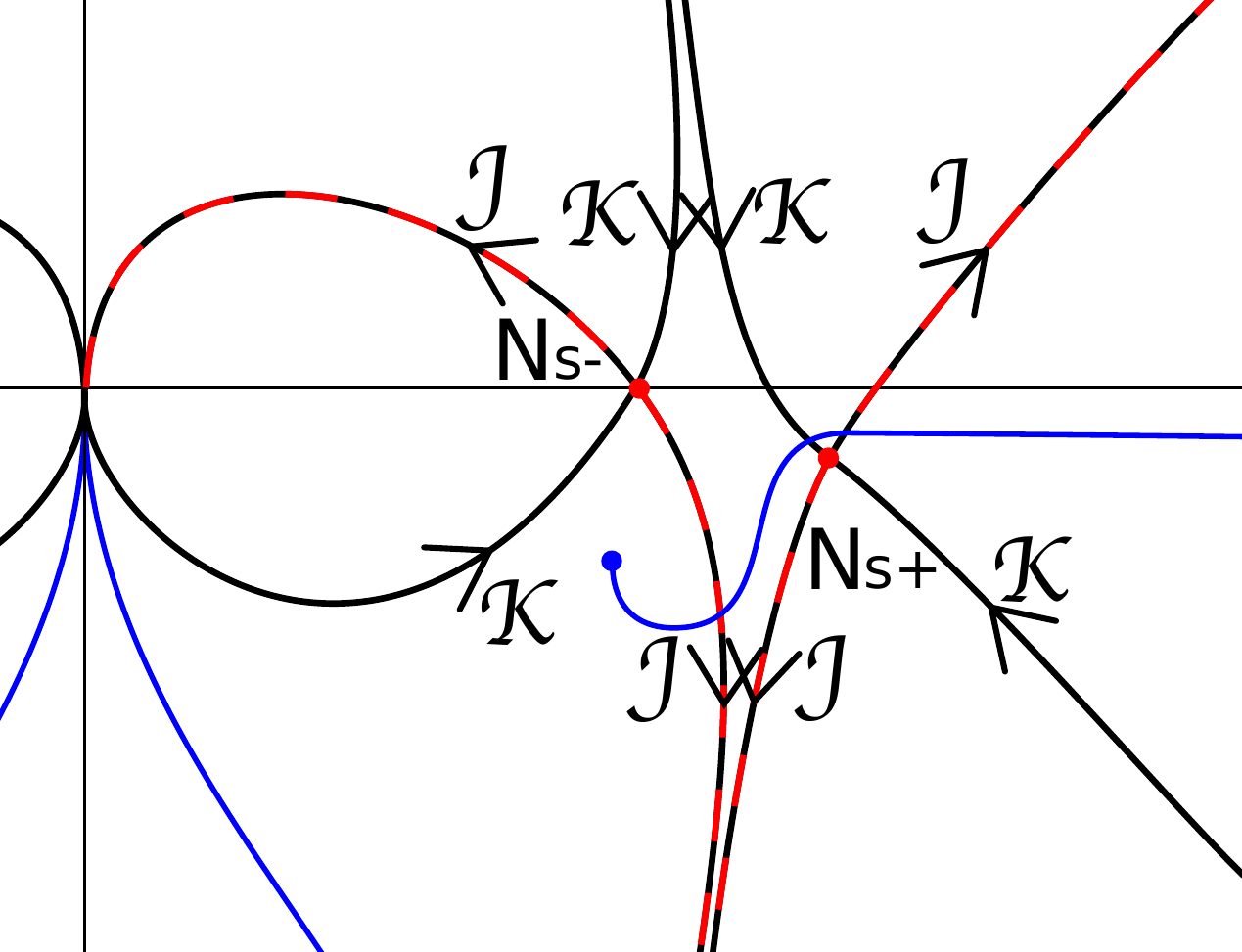} \hspace{.5cm}
\includegraphics[width=0.45\textwidth]{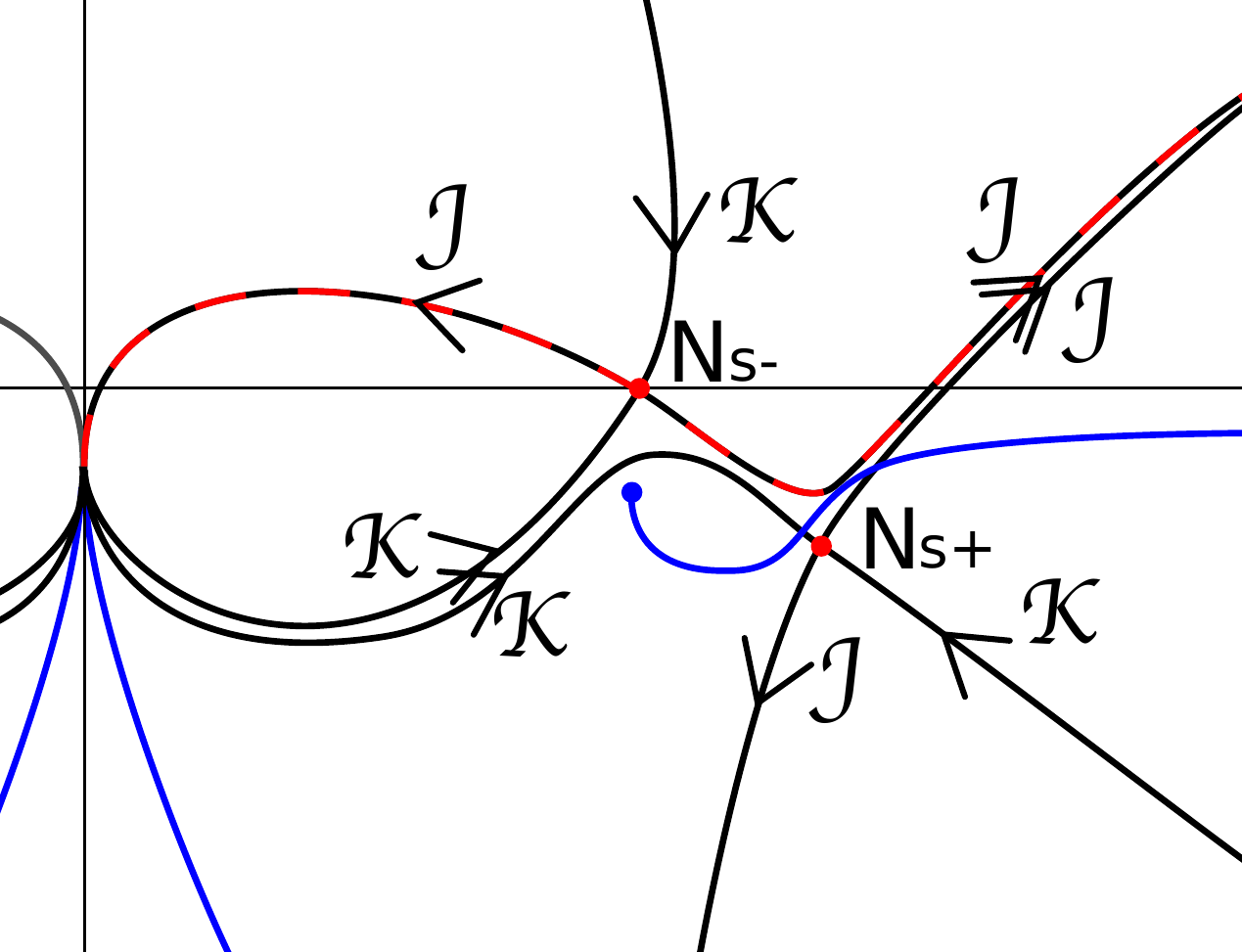}
\caption{{\it Left panel:} When the initial spread $\sigma$ is small, the relevant thimbles are forced to pass through the ``singular curve'' where the perturbative action blows up due to regions of vanishing scale factor in the geometries that are summed over.   {\it Right panel:} For a sufficiently large localization $\sigma \gtrsim \sigma_c,$ the relevant thimble stays above the singular curve, the path integral for the perturbations is well-defined and, with a judicious choice of initial state, stable fluctuations may be imposed.} \label{fig:SingularLines}
\end{figure}

However, selecting the appropriate saddle point solution is not enough. An important issue for the consistency of the calculation is that the perturbative action ought to be well-defined along the entire Lefschetz thimble that one integrates over in order to obtain the Feynman transition amplitude. An obstruction occurs whenever an (off-shell) background spacetime that is being integrated over contains a region where the scale factor of the universe reaches zero or passes through zero. At such locations the perturbations blow up, and the perturbative action $S^{(2)}$ becomes infinite, rendering the path integral ill-defined. Note that in such a case the surface form of the action \eqref{eq:boundary} would be augmented by one or more (infinite) surface terms at the intermediate times where $\bar{q}=0.$ Thus, in order for the path integral for background and fluctuations to be well-defined, we must ensure that the whole Lefschetz thimble does not include any backgrounds containing regions where $\bar{q}(t)=0,$ for $0 < t < 1.$ The occurrence of singular regions depends on the value of the localization $\sigma,$ according to Eq. \eqref{qsol2}. As discussed in subsection \ref{subsec:fluctuations}, when $\sigma=0$ the locus of singular backgrounds consists of the parts of the real $N$ line with $|N| \geq N_\star.$ Thus, with vanishing localization we not only obtain interference between two background solutions, but the thimbles also cross a region where the perturbative action is ill-defined since $N_{s+}>N_\star$. For small non-zero localization $\sigma,$ the singular curve moves away from the real $N$ line into the lower right quadrant in the complex $N$ plane, see the left panel of Fig. \ref{fig:SingularLines}. At this point, the thimbles still cross the singular curve. As the localization is further increased, we find that there exists a critical value which, as we will shortly see, is closely related to (though slightly larger than) the critical value $\sigma_c$ for the Stokes phenomenon. At this value the single remaining thimble just touches the singular curve of infinite $S^{(2)},$ and beyond this value of the localization the thimble is everywhere well-defined. This configuration is illustrated in the right panel in Fig. \ref{fig:SingularLines}.

In order to determine the critical value of the localization quantitatively, we will consider the physically relevant limit in which the final scale factor $a_1 = \sqrt{q_1}$ becomes large. In this limit we are able to obtain an analytic description of the thimble. Then we shall work out the locus of the singular curve consisting of all points in the complex $N$-plane for which $q(t)$ has a zero for some $0<t<1.$ This will allow us to find the intersections of the thimble with the singular curve, as a function of the localization $\sigma$. Since the algebra is a little involved, it is helpful to scale out dimensions. We therefore set $\Lambda=3 H^2 M_{Pl}^2$, $N=\tilde{N} H^{-2}$, $q=\tilde{q} H^{-2}$, $a=\tilde{a} H^{-1}$, $p=-3 \tilde{a}_i V_3$ (the classical value corresponding to an expanding universe), $\tilde{\hbar}=\hbar H^2/(M_{Pl}^2 V_3)$,  and $\sigma=H^{-2} \tilde{a}_i^{1\over 2} s \sqrt{\tilde{\hbar}}$. It is also helpful to define $\tilde{a}_1 = \gamma \tilde{a}_i$ and $\tilde{N}=\tilde{a_i} n$. We now calculate the total exponent minus that for the saddle point $N_{s -}$, in the limit of large $\gamma$, obtaining 
\begin{align}
& i\bar{S}^{(0)}+ip(q_i)\bar{q}_0-\frac{\hbar(\bar{q}_0-q_i)^2}{4\sigma^2} - \left[i\bar{S}^{(0)}+ip(q_i)\bar{q}_0-\frac{\hbar(\bar{q}_0-q_i)^2}{4\sigma^2}\right]_{N_{s-}} \nonumber \\ \approx \, &  i a_i^3(\delta n^2(-3 +9i s^2)+\delta n^3 )/\tilde{\hbar} +O(1/\gamma),
\label{acteq}
\end{align}
where we have defined $\delta n\equiv n-(\gamma-1)$ as the deviation of $n$ from its saddle point value. Requiring that the imaginary part of (\ref{acteq}) is zero is the defining condition for the thimble, and provides a relation between the real and imaginary parts of $\delta n$, which in turn determines the locus of the thimble in the complex $n-$plane. 

It follows from (\ref{qsol2}) that the condition for $q(t)$ to possess a zero for $0<t<1$ is that $\alpha$ or $\beta$, given in (\ref{eq:betaGen}), are real and lie between zero and one, at the given value of $N$.  But $\alpha$ and $\beta$ are just the two roots $g$ of a quadratic equation,
\begin{equation}
g^2+{\gamma^2-1-n^2+6 i s^2 \over n(n+3 i s^2) }g +{n+3 i s^2(\gamma^2-2 n -n^2)\over n^2(n+3 i s^2)}=0.
\label{gameq}
\end{equation}
Since we are interested in values of $n$ for which $g$ is real and lies between zero and one, we can set the imaginary part of (\ref{gameq}) to zero and obtain $g$ in terms of $n$. Substituting this expression for $g$ into the real part of (\ref{gameq}) gives a relation between the real and imaginary parts of $n$ which determines the locus of the points in the complex $n-$plane for which $q(t)$ vanishes for some $0<t<1$. (One must also check that indeed $g$ lies between zero and one).  Setting $n=\gamma-1+\delta n$ as above, and again taking the limit of large $\gamma$, we find a relation between the real and imaginary parts of $\delta n = x+i y$: 
\begin{equation}
6 s^2 y^3+y^2+6 s^2 y(x^2-x)+9 x^2s^4=0.
\label{gameq1}
\end{equation}
In terms of the same variables, it follows from setting zero the imaginary part of the exponent in Eq. \eqref{acteq} to its value at the saddle that the locus of the thimble is given by
\begin{equation}
-3 x^2 + x^3-18 s^2 x y+3 y^2-3 x y^2=0.
\label{thimblelocus}
\end{equation}
The smaller root $y=(9 s^2 x-\sqrt{3}\sqrt{3 x^2+27 s^4 x^2-4 x^3+x^4})/(3(1-x))$ of (\ref{thimblelocus})  corresponds to the right hand portion of the thimble. Substituting this back into the left hand side of (\ref{gameq1}), we plot the result against $x$ for various values of $s$. At low values of $s$ there are two nonzero roots which move towards each other as $s$ is increased. There is a critical value of $s$ which corresponds to the thimble just touching the line of zeroes of $q(t)$. Above this value there is no real, nonzero root. We find the critical value  $s_c \approx 0.46937$. Restoring the units in which our bound (\ref{stokesbound}) is expressed, we are able to conclude that the Lefschetz thimble relevant to the background solution encounters no zero in $q(t)$, and hence the stable mode persists across the entire relevant Lefschetz thimble, provided that
\begin{align}
a_i^3 V_3 \gtrsim C \frac{\hbar}{M_{Pl} \sqrt{\Lambda}},
\label{moreaccbound}
\end{align}
with the constant $C\approx 0.382$. This bound represents a modest strengthening of the Stokes bound given in (\ref{stokesbound}) and characterizes the parameter range over which the path integral for the perturbations is well-defined, once the background has been integrated out.

\section{Discussion} \label{sec:discussion}

When inflation was discovered, there were hopes that it would explain the initial conditions of the universe. In fact, it was thought that inflation, being an attractor, would explain the starting point of the hot big bang model irrespective of what came before it. According to this view,  if inflation had taken place, we would never have to, nor would we ever be able to, understand what occurred at earlier times. In recent years it has become increasingly clear that, at the classical level, the situation is more nuanced: inflation requires special initial conditions to get underway, and it is still actively debated whether such initial conditions are likely or unlikely (see, {\it e.g.}, \cite{Gibbons:2006pa,Ijjas:2013vea}, and the recent numerical works \cite{East:2015ggf,Clough:2016ymm,Marsh:2018fsu} which, however, neglect the crucial effect of the initial scalar field velocity). 

Our work in this paper adds an additional, purely quantum consideration: if inflation were truly  independent of what came before it, then we ought to be able to consider inflation all the way back to the singularity when the size of the universe approached zero.  And we ought to be able to assume that there was nothing prior to inflation. As we have shown, the de Sitter propagator from an initial vanishingly small three-geometry to a large final three-geometry does indeed become independent of the initial fluctuations, and provides an answer depending solely on the final fluctuations. At first sight, this would seem to reinforce the hopes described above, and might even suggest that at a quantum level inflation really can stand on its own and explain the state of the universe. However, our calculations reveal that, semi-classically at least, there is a tension between the background and  the fluctuations, and the wrong-sign Bunch-Davies mode functions are selected for the fluctuations. This result immediately implies a breakdown of the model, {\it i.e.}, that one cannot describe the origin of an inflationary universe of vanishingly small initial size. Our calculation demonstrates that quantum gravity effects cannot be ignored at the beginning of inflation -- put differently, the beginning of inflation is highly sensitive to UV effects, not just in the sense of its potential being sensitive to curvature corrections etc., but also in terms of its quantum vacuum. Since all predictions of inflation depend sensitively on the quantum vacuum, this is not a small issue.

Can one think of ways to avoid the negative result just described? An obvious possibility is to consider a pre-inflationary phase, perhaps a radiation phase before inflation. In fact, this was a popular idea in the early days of inflation. A pre-inflationary phase would have to set up the initial conditions for inflation, both classically (in terms of preparing a region of the universe that is sufficiently flat and at a sufficiently high and homogeneous energy density) and quantum mechanically (in terms of preparing the Bunch-Davies vacuum). We have provided the first steps in this direction by showing that quantum field theory in curved spacetime may be recovered as long as the spatial volume at the onset of inflation is large enough, more specifically as long as it is larger than 
\begin{align} 
a_i^3 V_3 \gtrsim \frac{\hbar \nu^2}{\sqrt{\Lambda}}\,,
\end{align}
assuming an expanding initial state. This result shows that the background may be treated classically from this size onwards, but the quantum state of the fluctuations remains unexplained by these arguments. Note that for a sub-Planckian Hubble constant during inflation, the size at which QFT in curved space-time breaks down is much larger than the Planck scale. A challenge for the future will be to investigate under what circumstances a pre-inflationary phase can also prepare the required Bunch-Davies state for the perturbations. 

We believe that these results change the status of inflation: instead of ``creating'' a flat universe, and ``creating'' the primordial fluctuations, inflation may better be thought of as a mechanism reinforcing flatness, and processing fluctuations out of a pre-existing (but assumed) quantum vacuum into classical density fluctuations. Inflation can exist as a phase of cosmological evolution, sandwiched between other phases, but it does not by itself explain the initial state. On the one hand, this significantly weakens our ability to test inflation observationally. After all, other types of cosmological evolution, such as ekpyrosis \cite{Lehners:2008vx}, can perform analogous processing tasks. On the other hand, we reach the highly welcome prospect that early universe cosmology offers us a window onto full quantum gravity, and not just QFT in curved spacetime. 


\acknowledgments

We would like to thank all of the participants of the workshop ``The path integral for gravity'', held at Perimeter Institute in November 2017 and, in particular, Angelika Fertig and Laura Sberna for stimulating discussions and collaboration with JF and NT on Ref.~\cite{Feldbrugge:2019}. ADT and JLL gratefully acknowledge the support of the European Research Council in the form of the ERC Consolidator Grant CoG 772295 ``Qosmology''. Research at Perimeter Institute is supported by the Government of Canada through Industry Canada and by the Province of Ontario through the Ministry of Economic Development, Job Creation and Trade.

\bibliographystyle{utphys}
\bibliography{QuantumIncompleteness}


\end{document}